\documentclass[a4paper,fleqn]{pazha2}

\usepackage{graphicx}
\usepackage{natbib}
\usepackage[english,russian]{babel}

\usepackage[T2A]{fontenc}
\usepackage{amsmath}
\usepackage{amsfonts}
\usepackage{amssymb}
\usepackage{epsf}
\usepackage{hyperref}
\usepackage{float}
\usepackage{fancyhdr}

\usepackage{amsmath}

\usepackage{caption}
\usepackage{multicol, blindtext}

\DeclareGraphicsExtensions{.pdf,.png,.jpg,.eps}

\usepackage{natbib}
\usepackage{array}
\newcolumntype{H}{>{\setbox0=\hbox\bgroup}c<{\egroup}@{}}

\usepackage[mathscr]{eucal}
\newcommand{\grs}{GRS~1739-278~}

\usepackage{multirow}
\begin{document}
\title{Investigation of the Outburst Activity
of the Black Hole Candidate GRS 1739-278}

\author{\bf  S. D. ~Bykov  \address{1,2}\email{sdbykov@edu.hse.ru}, E. V. ~Filippova \address{1}\email{sdbykov@edu.hse.ru,kate@iki.rssi.ru}, I. A. ~Mereminskiy \address{1}, A. N. ~Semena\address{1}, A. A.~ Lutovinov\address{1,2}\\
{\it
{$^1$}{Space Research Institute, Russian Academy of Sciences, Profsoyuznaya ul. 84/32, Moscow, 117997 Russia}\\
{$^2$}{Higher School of Economics, Myasnitskaya 20, 101000 Moscow, Russia}}}

\begin{abstract}
We have performed a joint spectral and timing analysis of the outburst of GRS 1739-278
in 2014 based on Swift and INTEGRAL data. We show that during this outburst the system exhibited
both intermediate states: hard and soft. Peaks of quasi-periodic oscillations (QPOs) in the frequency
range 0.1-5 Hz classified as type-C QPOs have been detected from the system. Using Swift/BAT data
we show that after the 2014 outburst the system passed to the regime of mini-outburst activity: apart from
the three mini-outbursts mentioned in the literature, we have detected four more mini-outbursts with a
comparable ($\sim$20 mCrab) flux in the hard energy band (15-50 keV). We have investigated the influence
of the accretion history on the outburst characteristics: the dependence of the peak flux in the hard energy
band in the low/hard state on the time interval between the current and previous peaks has been found (for
the outbursts during which the system passed to the high/soft state).
\end{abstract}

\section{INTRODUCTION}
During outbursts the transient systems with
black hole candidates exhibit several characteristic
states. The hardness-intensity and hardness-rms
diagrams, on which the systems, as a rule, exhibit
characteristic dependences, are used for their
classification (Grebenev et al. 1993; Tanaka and
Shibazaki 1996; Remillard and McClintock 2006;
Belloni 2010; Belloni and Motta 2016). Initially,
two states of such systems were detected: low/hard
and high/soft (see Remillard and McClintock (2006)
and references therein). According to the most
popular accretion flow model, in such systems it is
believed that the accretion disk in the low/hard state
is truncated at a large inner radius, while the region
between the disk and the compact object is filled with
an optically thin hot plasma, a corona, which makes
a major contribution to the source's emission in the
form of a power law with a high-energy cutoff. As the
outburst develops, the inner radius of the accretion
disk decreases and in the high/soft state the accretion
disk makes a major contribution to the system's
emission (Grebenev et al. 1997; Gilfanov 2010).

Subsequently, it was found that between these
well-defined states the system could be in transition ones, whose established classification is currently absent
(Remillard and McClintock 2006; Belloni and
Motta 2016). In this paper we used the classification
from Belloni and Motta (2016), according to which
the source exhibits hard and soft intermediate states.
These states are characterized by both thermal and
power-law components in the source's energy spectrum,
but they do not differ greatly from the viewpoint
of their spectral characteristics. Nevertheless, there
are several distinctive features of these states. In the
hard intermediate (and low/hard) state peaks of type C quasi-periodic oscillations (QPOs) and broadband
noise at low frequencies with a fractional rms of both
components of tens of percent are often observed in
the source's power spectrum. Type-B QPOs, which,
as a rule, have a fractional rms of a few percent,
and weak (rms < 10\%), frequency-dependent noise
at low frequencies are detected in the soft intermediate
state (Belloni and Motta 2016). The QPO
frequency has inverse (type C) and direct (type B)
dependences on the flux in the power-law component
(Motta et al. 2011). The most popular QPO model
is based on the Lense-Thirring precession of a hot
corona near the compact object (Ingram et al. 2009),
but there is no full physical picture for the formation
of QPOs of different types. In the low/hard and
hard intermediate states the systems with black hole
candidates are observed in the radio band. During
the transition to the soft intermediate state the system crosses the so-called "jet line"$\,$ and ceases to radiate
in the radio band (Remillard and McClintock 2006;
Belloni 2010). Thus, an investigation of the intermediate
states is required for a more detailed study of
the physical processes responsible for the formation
of QPOs and jets.

It is believed that during an outburst the system
must pass from the low/hard to high/soft state
and back, exhibiting the intermediate states and a
characteristic "q"$\,$ shape on the hardness-intensity
diagram (Belloni and Motta 2016). However, many
sources with black hole candidates exhibit the so-called
failed outbursts, when the system does not
reach the high/soft state (Ferrigno et al. 2012, 2014a;
Del Santo et al. 2015; Mereminskiy et al. 2017).

The same system can exhibit both types of outbursts
(Motta et al. 2010; Furst et al. 2015; Mereminskiy
et al. 2017), with the same type of outbursts
occurring at peak luminosities differing by tens of
times. For example, in GRS 1739-278 the transition
to the high/soft state was observed both during bright
outbursts, when the peak flux reached 300 mCrab
(15-50 keV), and mini-outbursts with a peak flux of
40 mCrab (15-50 keV) (Yan and Yu 2017). Different
types of outbursts occur at close peak luminosities -
a failed outburst at a peak flux of 30 mCrab in the 15-
50 keV energy band was detected in the same system
GRS 1739-278 (Mereminskiy et al. 2017).

At present, there is no full understanding of what
physical conditions are necessary for the system's
transition from the low/hard to high/soft state, but
it is clear that not only the change in accretion rate
is responsible for this process. A hysteresis behavior
is observed on the hardness-intensity diagram
even within one full-fledged outburst of the source,
i.e., the transition from the hard state to the soft
one occurs at luminosities greater than the luminosity
during the transition from the soft state to the
hard one by several (or even tens of) times. The
corona size (Homan et al. 2001), the accretion disk
size (Smith et al. 2001), the evolution history of the
inner accretion disk radius (Zdziarski et al. 2004),
and the accretion disk mass (Yu et al. 2004; Yu and
Yan 2009) are considered as additional parameters.
Yu et al. (2007) and Wu et al. (2010) found a correlation
between the luminosity at which the transition
from the low/hard to high/soft state occurs and the
peak luminosity in the high/soft state, based on which
they put forward the idea about the influence of the
accretion disk mass on the outburst evolution, and
a correlation between the peak flux in the low/hard
state and the time interval between the current and
previous peak fluxes in the low/hard state. Based
on these dependences, the authors hypothesized that
the peak luminosity in the low/hard state is also determined by the mass of the accretion disk accumulated
between outbursts. Yu et al. (2004, 2009)
used mostly low-mass systems with neutron stars,
which accounted for about 70\% of the samples, to
construct the relationship between the luminosity of
the transition from the low/hard to high/soft state
and the peak luminosity in the high/soft state. In
contrast, the correlation between the peak flux in
the low/hard state and the time interval between the
current and previous peak fluxes in the low/hard state
was found only for one system, GX~339-4 (based
on eight outbursts detected from 1991 to 2006), and
requires a further confirmation based on data for the
outbursts from 2006 to 2018. Thus, an investigation
of transient systems exhibiting several outbursts is
required to develop a physical model for the outbursts
of binary systems with black hole candidates, which
allows the evolution of binary parameters from outburst
to outburst to be measured. GRS 1739-278
belongs to such sources.

\subsection{\grs}
The X-ray source GRS 1739-278 was discovered
by the SIGMA coded-mask telescope onboard the
GRANAT observatory on March 18, 1996 (Paul
et al. 1996). The peak flux from the source during
the first recorded outburst was $\sim 800 - 1000$ mCrab in
the 2-10 keV energy band (Borozdin et al. 1998).
The corresponding radio source was detected by
Durouchoux et al. (1996). During the 1996 outburst
the system passed from the low/hard to high/soft
state (Borozdin et al. 1998). QPO peaks at 5 Hz
were detected in the source (Borozdin and Trudolyubov
2000). From the shape of the power spectrum
and the total fractional rms it can be concluded
that the system was recorded in the soft intermediate
state and the recorded QPOs are type-B ones.

The second outburst was recorded by the
Swift/BAT monitor on March 9, 2014 (Krimm
et al. 2014). A spectral analysis of this outburst
based on Swift/XRT data was performed by Yu and
Yan (2017) and Wang et al. (2018). They showed
that during the outburst the system passed from the
low/hard to high/soft state through the intermediate
one, but no detailed analysis of the intermediate state
was made. At the outburst onset (March 19, 2014)
the system was also recorded by the INTEGRAL
observatory. According to the analysis of these data,
the source was recorded up to energies of $\sim 200$ keV,
while the energy spectrum was fitted by a power law
with a cutoff with the following parameters: a photon
index $\Gamma=1.4\pm 2$ and a cutoff energy $E_{cut}=90^{+40}_{-20}$ keV (Filippova et al. 2014b). The source was observed
by the NuSTAR observatory (Miller et al. 2015) 17 days after the outburst onset (March 26, 2014).
A spectral analysis of the NuSTAR data showed that
the system continued to be in the low/hard state. The
reflected component of the hard emission presumably
associated with an accretion disk whose inner radius
must reach the innermost stable orbit was observed
in the source's spectrum, but the disk component
itself was not recorded in the spectrum. Mereminskiy
et al. (2019) performed a detailed timing analysis of
the NuSTAR data, based on which QPOs at frequencies
0.3-0.7 Hz were detected in the variability power
spectrum for GRS 1739-278. Based on data from the
MAXI monitor, Wang et al. (2018) showed that the
system passed back from the high/soft to low/hard
state in November-December 2014.

Two mini-outbursts (the peak flux in the 15-50 keV energy band was $\sim 40$ mCrab) were detected
from the system 200 days after this outburst. A
spectral analysis of the Swift/XRT data showed that
during these mini-outbursts the system passed from
the low/hard to high/soft state and returned back to
the low/hard one (Yu and Yan 2017).

The next outburst from the source was detected in
September 2016 (Mereminskiy et al. 2016). During
this outburst the peak flux in the 20-60 keV energy
band was $\sim 30$ mCrab. Based on our analysis of the
INTEGRAL and Swift data, it was shown that during
the outburst the system was in the low/hard state
and exhibited no transition to the high state. i.e., the
outburst turned out to be a failed one (Mereminskiy
et al. 2017). Some softening of the spectrum was
recorded, i.e., the photon index increased from 1.73
at the outburst onset to 1.86 at the peak flux, with
the power law having been observed up to energies
$\sim 150$ keV without cutoffs; no contribution to the
emission from the accretion disk was recorded.

In this paper for the first time we have performed
a simultaneous study of the spectral evolution and
temporal variability of the system over the entire 2014
outburst, which has allowed the source's intermediate
states to be classified in detail, made a comparative
analysis of the system's behavior in all of the outbursts
detected to date, and investigated the system's
behavior between outbursts.

\vspace{-0.1cm}
\section{OBSERVATIONS}
\subsection{Swift Data}
GRS 1739-278 during the 2014 outburst was
observed by the Swift/XRT telescope in the windowed
timing mode (Burrows et al. 2005) from
March 20, 2014 (MJD 56736) to November 1, 2014
(MJD 56962), i.e., the observations were begun
on the 11th day after the outburst onset. A total
of 39 observations with ObsID 000332030XY were carried out; below we will use the last two digits to
denote the observation number.

The Swift/XRT light curves and spectra of the
source were obtained using an online repository
(Evans et al. 2007). Events with grade 0 were selected
to analyze the energy spectra. The source's spectra
were investigated in the 0.8-10 keV energy band,
because at energies below 0.8 keV the response matrix
is known inaccurately due to instrumental effects.
The spectra obtained through the online repository by
adding the counts in different bins were brought to
the form in which there were at least 100 counts per
energy channel with the grppha utility. This allowed
the $\chi^2$ statistic to be used in fitting the spectra in the
Xspec package (Arnaud 1996). Instrumental features
at energies of 1.8 and 2.3 keV were detected in several
spectra (marked in Table 1 by $\dagger$)\footnote{\url{https://heasarc.gsfc.nasa.gov/docs/heasarc/caldb/swift/docs/xrt/SWIFT-XRT-CALDB-09_v19.pdf}}. Therefore, when
fitting these data, we, following the recommendations
given at the mentioned link, used the gain command
in the Xspec package, which modifies the response
matrix by shifting the energies at which it was determined.
The offset parameters are given in Table 2.
Deviations of the data from the model at energies
below 1 keV were also observed in several spectra
(marked in Table 1 by $\star$). This feature of the spectrum
is related to the position of the source's image on
the telescope's detector\footnote{\url{https://heasarc.gsfc.nasa.gov/docs/heasarc/caldb/swift/docs/xrt/SWIFT-XRT-CALDB-09_v20.pdf}}. Using the telescope's
response matrix dependent on the source's position
on the detector (swxwt0s6psf1\_20131212v001.rmf)
allowed the quality of the data fit at low energies to be
improved.

To construct the light curves, we used the 0.5-10 keV energy band, events with grade 0-2, the
data were averaged over one observation. The typical
exposure time for XRT observations was $\sim 1-2$ ks
(see Table 1). The light curves used for our Fourier
analysis were constructed in the 0.5-10, 0.5-3, and
3-10 keV energy bands with a time resolution of
0.01 s.

The Swift/BAT light curve of the source was
retrieved from the online database of light curves
(Krimm et al. 2013).
\vspace{-0.2cm}
\subsection{INTEGRAL Data}

We also used the data from the JEM-X and
ISGRI/IBIS telescopes onboard the INTEGRAL
observatory (Winkler et al. 2003) processed with
the HEAVENS service (Walter et al. 2010). These data in combination with the quasi-simultaneous
Swift/XRT spectral data were used to construct
the source's broadband spectra in the energy range
0.8-200 keV. When fitting the spectra, we took
into account the systematic errors of 1 and 3\% for
the ISGRI\footnote{\url{https://www.isdc.unige.ch/integral/download/osa/doc/10.2/osa\_um\_ibis/index.html}} and JEM~-~X\footnote{\url{https://www.isdc.unige.ch/integral/download/osa/doc/10.2/osa\_um\_jemx/index.html}} instruments, respectively.
The times and exposures of observations for the
broadband spectra are given in Table 3.

\begin{table*}[]
\caption{Parameters of the best-fit models for the XRT energy spectra }
\label{tbl-xrt-1}
\hspace*{-1cm}\begin{tabular}{ccc|cc|cc|c|c|c}
$ID^a$                  &\begin{tabular}[c]{@{}c@{}}MJD-\\ -56000$^b$\end{tabular} & \begin{tabular}[c]{@{}c@{}}Exp. XRT$^c$\\ s\end{tabular}         & \begin{tabular}[c]{@{}c@{}}$N_H^d$,\\  $10^{22}$cm$^{-2}$\end{tabular}                  & $\Gamma^e$      & \begin{tabular}[c]{@{}c@{}}$T_{in}$, \\ keV $^f$\end{tabular}               &\begin{tabular}[c]{@{}c@{}}$R_{in} cos^{-1/2}(i)$, \\ km $^g$\end{tabular}                   & Flux$^h$               & $f_{po}^j$ & $\chi_N^2/dof$ \\ \hline\hline
$01$                  & 736                  & $79$                    & $1.51_{-0.63}^{+0.67}$ & $1.24_{-0.36}^{+0.37}$ & -                      & -                             & $1.57_{-0.29}^{+0.13}$ & 100  & $0.31/7$   \\ \hline
$02$                  & $741$                                                                                                                         & $2067$                  & $1.64\pm0.04$          & $1.38\pm0.03$ & -                      & -                          & $3.49\pm0.04$           &    100        & $1.09/459$   \\ \hline
\multirow{5}{*}{$03$} & \multirow{5}{*}{742} & \multirow{5}{*}{$1924$} & $1.65\pm0.03$          & $1.38\pm0.02$          & -                      & -                             & $4.53\pm0.04$          & 100 & $1.06/496$ \\
                     &                      &                         & $2.08_{-0.1}^{+0.11}$  & $1.5$                  & $0.24_{-0.03}^{+0.02}$ & $196.05_{-52.37}^{+70.17}$    & $4.44_{-0.04}^{+0.03}$ & $89\pm5$  & $1.11/495$ \\
                     &                      &                         & $2.96\pm0.09$          & $1.8$                  & $0.18\pm0.01$          & $1378.65_{-198.48}^{+235.78}$ & $4.2\pm0.03$           & $51\pm4$ & $1.64/495$ \\
\hline

\multirow{5}{*}{$04^{\dagger}$} & \multirow{5}{*}{747}                                                                                                          & \multirow{5}{*}{$2463$} & $1.82_{-0.06}^{+0.08}$ & $2.05_{-0.04}^{+0.08}$ & -                      & -                            & $7.29_{-0.08}^{+0.07}$  &   $100$    & $1.3/598$    \\
                      &                                                                                                                               &                        & $1.52_{-0.02}^{+0.08}$ & $1.5$                  & $0.79_{-0.06}^{+0.04}$ & $15.7_{-0.73}^{+3.94}$       & $7.87_{-0.11}^{+0.06}$  & $82\pm4$    & $1.28/597$ \\
                      &                                                                                                                               &                         & $1.74_{-0.06}^{+0.08}$ & $1.8$                  & $0.65_{-0.09}^{+0.05}$ & $19.22_{-3.96}^{+6.33}$      & $7.62_{-0.1}^{+0.03}$   &    $90\pm5$   & $1.27/597$   \\
                      &                                                                                                                               &                         & $1.9_{-0.07}^{+0.09}$  & $2$                    & $0.51_{-0.06}^{+0.05}$ & $25.76_{-10.15}^{+12.89}$    & $7.41_{-0.23}^{+0.03}$  &  $94\pm4$     & $1.28/597$    \\
\hline
                      \multirow{5}{*}{$05^{\dagger}$} & \multirow{5}{*}{751} & \multirow{5}{*}{$1832$} & $1.57_{-0.07}^{+0.08}$ & $1.88\pm0.05$          & -                      & -                            & $6.76_{-0.05}^{+0.06}$  &    100   & $1.17/552$    \\
                     &                      &                         & $1.37\pm0.06$          & $1.5$                  & $0.76_{-0.06}^{+0.1}$  & $12.72_{-1.51}^{+1.98}$      & $7.17_{-0.14}^{+0.05}$  & $88\pm5$    & $1.18/551$  \\
                     &                      &                         & $1.62_{-0.1}^{+0.07}$  & $1.8$                  & $0.55_{-0.1}^{+0.15}$  & $18.49_{-6.13}^{+13.25}$     & $6.87_{-0.2}^{+0.03}$   &     $96\pm4$  & $1.17/551$ \\
                     &                      &                         & $1.85_{-0.07}^{+0.1}$  & $2$                    & $0.26_{-0.03}^{+0.1}$  & $143.63_{-71.44}^{+135.22}$  & $6.62_{-0.08}^{+0.04}$  &     $93\pm7$  & $1.18/551$ \\

 \hline
\multirow{4}{*}{$06$} & \multirow{4}{*}{757}                                                                                                          & \multirow{4}{*}{$1811$} & $2.26\pm0.03$          & $2.25\pm0.02$ & -                      & -                          & $10.67_{-0.08}^{+0.07}$ & 100        & $1.43/496$   \\
                      &                                                                                                                               &                         &$1.74\pm0.03$          & $1.5$    & $1.12\pm0.04$          & $14.24_{-0.85}^{+0.94}$    & $10.56_{-0.11}^{+0.07}$ & $53\pm4$ & $1.28/495$               \\
                      &                                                                                                                               &                         &$1.85\pm0.03$          & $1.8$    & $1.13\pm0.05$          & $12.09_{-0.84}^{+0.97}$    & $10.51\pm0.1$           & $66\pm6$ & $1.25/495$               \\
                      &                                                                                                                               &                         & $1.95\pm0.03$          & $2$           & $1.21_{-0.07}^{+0.06}$ & $9.21_{-0.68}^{+0.81}$     & $10.46_{-0.10}^{+0.09}$  & $75\pm8$         & $1.25/495$   \\
\hline
\multirow{5}{*}{$07$} & \multirow{5}{*}{761} & \multirow{5}{*}{$1612$} & $2.15\pm0.03$          & $2.11\pm0.02$          & -                      & -                             & $8.35_{-0.05}^{+0.06}$ &  & $1.53/514$ \\
                     &                      &                         & $2.15_{-0.04}^{+0.05}$ & $1.5$                  & $0.61\pm0.02$          & $41.96_{-4.21}^{+4.7}$        & $8.9_{-0.1}^{+0.05}$   & $70\pm2$ & $1.13/513$ \\
                     &                      &                         & $2.31_{-0.05}^{+0.06}$ & $1.8$                  & $0.49\pm0.02$          & $62.57_{-9.27}^{+10.74}$      & $8.68_{-0.08}^{+0.05}$ & $78\pm3$  & $1.1/513$  \\
                     &                      &                         & $2.43\pm0.07$          & $2$                    & $0.4_{-0.02}^{+0.03}$  & $97.64_{-20.61}^{+24.37}$     & $8.47_{-0.1}^{+0.03}$  & $82\pm4$ & $1.21/513$ \\
\hline

\multirow{4}{*}{$08^{\star}$}                  & \multirow{4}{*}{764}                                                                                                                    & \multirow{4}{*}{$808$}                   & $2.27\pm0.05$          & $2.28\pm0.03$ & -                      & -                          & $10.11_{-0.10}^{+0.11}$  & 1  00      & $1.36/366$   \\
                      &                                                                                                                               &                         & $1.69\pm0.04$          & $1.5$    & $1.17\pm0.06$          & $13.15_{-1.07}^{+1.23}$    & $9.91_{-0.2}^{+0.11}$   & $48\pm7$ & $1.16/365$               \\
                      &                                                                                                                               &                         & $1.79\pm0.05$          & $1.8$    & $1.2\pm0.07$           & $11.27_{-0.94}^{+1.15}$    & $9.86_{-0.21}^{+0.16}$  & $60\pm10$ & $1.16/365$               \\ 
                      &                                                                                                                               &                         & $1.89\pm0.05$          & $2$           & $1.28_{-0.09}^{+0.08}$ & $9.13_{-0.73}^{+0.89}$     & $9.80_{-0.14}^{+0.17}$   & $67\pm12$         & $1.16/365$   \\
\hline
\multirow{5}{*}{$09^{\dagger}$} & \multirow{5}{*}{771} & \multirow{5}{*}{$1872$} & $1.75_{-0.11}^{+0.1}$  & $2.08\pm0.08$          & -                      & -                            & $6.14_{-0.11}^{+0.07}$  &  100     & $1.14/504$\\
                     &                      &                         & $1.4_{-0.09}^{+0.07}$  & $1.5$                  & $0.84_{-0.07}^{+0.12}$ & $12.87_{-2.89}^{+2.9}$       & $6.57_{-0.14}^{+0.09}$  & $81\pm7$    & $1.14/503$ \\
                     &                      &                         & $1.58_{-0.09}^{+0.1}$  & $1.8$                  & $0.73_{-0.1}^{+0.16}$  & $12.53_{-4.37}^{+6.32}$      & $6.39_{-0.18}^{+0.07}$  &   $90\pm8$    & $1.14/503$ \\
                     &                      &                         & $1.74_{-0.08}^{+0.1}$  & $2$                    & $0.63_{-0.15}^{+0.29}$ & $11.57_{-8.42}^{+13.53}$     & $6.21_{-0.19}^{+0.04}$  &      $96\pm7$ & $1.14/503$ \\
\hline
 \multirow{4}{*}{$10^{\star}$}                  & \multirow{4}{*}{776}                                                                                                                           & \multirow{4}{*}{$1507$}                  & $2.73\pm0.04$          & $2.22\pm0.02$ & -                      & -                             & $16.94_{-0.12}^{+0.1}$  & 100       & $1.37/525$  \\
                      &                                                                                                                               &                         & $2.32\pm0.04$          & $2$           & $1.42\pm0.06$          & $9.53_{-0.52}^{+0.57}$        & $16.45_{-0.14}^{+0.16}$ & $68\pm9$  & $1.09/524$ \\
                      &                                                                                                                               &                         & $2.51_{-0.06}^{+0.05}$ & $2.4$         & $1.85_{-0.05}^{+0.06}$ & $6.05_{-0.6}^{+0.58}$         & $16.26_{-0.26}^{+0.1}$  & $66\pm9$  & $1.12/524$ \\
                      &                                                                                                                               &                         & $1.8\pm0.02$           & -             & $1.72\pm0.02$          & $10.09_{-0.23}^{+0.24}$       & $15.53_{-0.13}^{+0.09}$ & 0         & $1.64/525$ \\ \hline
\multirow{4}{*}{$11^{\star \dagger}$}                  & \multirow{4}{*}{781}                                                                                                                           & \multirow{4}{*}{$1922$}               & $2.37_{-0.09}^{+0.13}$ & $1.72_{-0.03}^{+0.08}$ & -                      & -                       & $27.3_{-0.41}^{+0.34}$  &  100     & $1.74/525$ \\ 
                      &                                                                                                                               &                          & $1.92_{-0.05}^{+0.09}$ & $2$                    & $1.45_{-0.06}^{+0.03}$ & $15.95_{-1.49}^{+1.6}$  & $19.76_{-0.3}^{+0.22}$  & $11\pm7$    & $1.11/524$  \\
                                            &                                                                                                                               &                          &$2.02_{-0.18}^{+0.11}$ & $2.4$                  & $1.46_{-0.08}^{+0.07}$ & $16.07_{-1.91}^{+1.39}$      & $19.63_{-0.32}^{+0.27}$ &  $11\pm8$     & $1.11/524$  \\
                      &                                                                                                                               &                         & $1.81_{-0.06}^{+0.12}$ & -                      & $1.49_{-0.07}^{+0.05}$ & $15.58_{-1.24}^{+1.87}$ & $19.55_{-0.76}^{+0.18}$ &   0    & $1.12/525$    \\ \hline
\multirow{4}{*}{$12$}                  & \multirow{4}{*}{786}                                                                                                                           & \multirow{4}{*}{$1868$}                  & $2.38\pm0.03$          & $2.3\pm0.02$  & -                      & -                          & $8.64_{-0.05}^{+0.06}$  & 100        & $1.18/495$   \\
                      &                                                                                                                               &                         & $2.04\pm0.03$          & $2$           & $1.14\pm0.06$          & $9.73_{-0.72}^{+0.86}$     & $8.49\pm0.08$           & $74\pm6$         & $0.99/494$   \\
                      &                                                                                                                               &                         & $2.25\pm0.04$          & $2.4$         & $1.70_{-0.06}^{+0.08}$  & $4.2_{-0.58}^{+0.54}$      & $8.36_{-0.13}^{+0.04}$  & $77\pm9$         & $1.04/494$   \\
                      &                                                                                                                               &                         & $1.53\pm0.02$          & -             & $1.58\pm0.02$          & $8.4_{-0.19}^{+0.2}$       & $7.83_{-0.06}^{+0.05}$  & 0          & $1.97/495$   

\end{tabular}

\end{table*}

\begin{table*}[]
\caption*{Table 1 (Cont.)}
\hspace*{-0.5cm}\begin{tabular}{ccc|cc|cc|c|c|c}

$ID^a$                  &\begin{tabular}[c]{@{}c@{}}MJD-\\ -56000$^b$\end{tabular} & \begin{tabular}[c]{@{}c@{}}Exp. XRT$^c$\\ s\end{tabular}         & \begin{tabular}[c]{@{}c@{}}$N_H^d$,\\  $10^{22}$cm$^{-2}$\end{tabular}                  & $\Gamma^e$      & \begin{tabular}[c]{@{}c@{}}$T_{in}$, \\ keV $^f$\end{tabular}               &\begin{tabular}[c]{@{}c@{}}$R_{in} cos^{-1/2}(i)$, \\ km $^g$\end{tabular}                   & Flux$^h$               & $f_{po}^j$ & $\chi_N^2/dof$ \\ \hline\hline

\multirow{4}{*}{$13^{\star}$}                  & \multirow{4}{*}{791}                                                                                                                           & \multirow{4}{*}{$1887$}                  & $2.34\pm0.03$          & $2.34\pm0.02$ & -                      & -                             & $7.67_{-0.07}^{+0.05}$  & 100       & $1.16/497$ \\

                     &                                                                                                                               &                         &  $1.96\pm0.03$          & $2$           & $1.13\pm0.05$          & $9.76_{-0.66}^{+0.77}$        & $7.55_{-0.08}^{+0.07}$  & $71\pm6$  & $0.99/496$ \\
                     &                                                                                                                               &                         &  $2.12\pm0.04$                 & $2.4$         & $1.58\pm0.05$                 & $4.88^{+0.47}_{-0.50}$                        & $7.37^{+0.04}_{-0.08}$                  & $78\pm9$  & $1.02/496$ \\
                     &                                                                                                                               &                         &  $1.48\pm0.02$          & -             & $1.55\pm0.02$          & $8.27\pm0.19$                 & $6.96_{-0.07}^{+0.04}$  & 0         & $1.99/497$ \\\hline

\multirow{4}{*}{$14^{\star \dagger}$}                  & \multirow{4}{*}{796}                                                                                                                           & \multirow{4}{*}{$2098$}              &    $2.85_{-0.07}^{+0.09}$ & $2.27_{-0.03}^{+0.11}$ & -                      & -                       & $9.67_{-0.09}^{+0.07}$  &    100   & $1.43/542$  \\
                      &                                                                                                                               &                         & $2.09_{-0.08}^{+0.07}$ & $2$                    & $1.25\pm0.05$          & $11.74_{-1.62}^{+0.64}$ & $8.31_{-0.16}^{+0.07}$  & $43\pm10$    & $1.09/541$  \\
                      &                                                                                                                               &                         & $2.34\pm0.09$          & $2.4$                  & $1.28\pm0.07$          & $10.65_{-0.99}^{+1.59}$ & $8.17_{-0.14}^{+0.08}$  &   $54\pm10$    & $1.09/541$   \\
                      &                                                                                                                               &                         & $1.59_{-0.01}^{+0.09}$ & -                      & $1.42_{-0.04}^{+0.06}$ & $10.68_{-0.82}^{+0.79}$ & $7.66\pm0.07$           &  0     & $1.33/542$\\ \hline
\multirow{4}{*}{$15$}                  & \multirow{4}{*}{801}                                                                                                                           & \multirow{4}{*}{$1988$}                  & $2.46\pm0.03$          & $2.39\pm0.02$ & -                      & -                          & $8.68_{-0.06}^{+0.07}$  & 100        & $1.34/472$   \\
                      &                                                                                                                               &                         & $2.01\pm0.03$          & $2$           & $1.16\pm0.05$          & $10.86_{-0.65}^{+0.75}$    & $8.48_{-0.10}^{+0.08}$   & $65\pm6$         & $1.06/471$   \\
                      &                                                                                                                               &                         & $2.22\pm0.05$          & $2.4$         & $1.48_{-0.04}^{+0.05}$ & $6.06_{-0.56}^{+0.53}$     & $8.36_{-0.12}^{+0.04}$  & $73\pm10$         & $1.08/471$   \\
                      &                                                                                                                               &                         & $1.57\pm0.02$          & -             & $1.52\pm0.02$          & $9.31_{-0.22}^{+0.23}$     & $7.87_{-0.06}^{+0.05}$  & 0          & $1.81/472$  \\\hline

\multirow{4}{*}{$16^{\star}$} & \multirow{4}{*}{806}                                                  & \multirow{4}{*}{$1889$}    & $2.29\pm0.03$          & $2.35\pm0.02$ & -                      & -                             & $6.74_{-0.05}^{+0.04}$  & 100       & $1.12/524$ \\
 & &  & $2.01\pm0.03$          & $2$           & $0.92\pm0.05$          & $12.82_{-1.13}^{+1.33}$       & $6.76_{-0.06}^{+0.04}$  & $78\pm4$  & $1.05/523$ \\
& & & $2.25\pm0.03$          & $2.4$         & $1.72_{-0.14}^{+0.24}$ & $2.28_{-0.78}^{+0.69}$        & $6.65_{-0.16}^{+0.03}$  & $92\pm8$  & $1.11/523$ \\
& & & $1.43$                 & -             & $1.52$                 & $8.08$                        & $6.05$                  & 0         & $3.08/524$ \\\hline
\multirow{4}{*}{$17^{\star \dagger}$} & \multirow{4}{*}{811}                                                  & \multirow{4}{*}{$2045$}    & $3.13_{-0.39}^{+0.15}$ & $2.29_{-0.03}^{+0.05}$ & -                      & -                       & $10.6\pm0.12$           &    100   & $1.77/504$   \\
     &                                                      &           & $2.0_{-0.11}^{+0.07}$  & $2$                 & $1.25_{-0.04}^{+0.08}$ & $13.28_{-1.55}^{+1.06}$ & $8.69_{-0.26}^{+0.16}$  & $29\pm7$ & $1.11/503$ \\
     &                                                      &           & $2.19_{-0.12}^{+0.07}$ & $2.4$                  & $1.26_{-0.05}^{+0.06}$ & $12.72_{-1.26}^{+1.28}$      & $8.61_{-0.13}^{+0.08}$  &   $37\pm13$    & $1.12/503$    \\
     &                                                      &           & $1.66_{-0.02}^{+0.09}$ & -                      & $1.42_{-0.07}^{+0.04}$ & $11.31_{-0.67}^{+1.27}$ & $8.37_{-0.32}^{+0.04}$  &  0     & $1.23/504$   \\ \hline

\multirow{3}{*}{$19^{\star}$} & \multirow{3}{*}{823}                                                  & \multirow{3}{*}{$2355$}   &  $2.0\pm0.03$           & $2$           & $1.29\pm0.03$          & $9.94_{-0.3}^{+0.32}$         & $6.7_{-0.04}^{+0.06}$   & $41\pm5$  & $1.07/519$ \\
&&& $2.15\pm0.05$          & $2.4$         & $1.4\pm0.02$           & $8.23\pm0.28$                 & $6.64_{-0.06}^{+0.05}$  & $47\pm7$  & $1.12/519$ \\
&&& $1.72\pm0.02$          & -             & $1.46\pm0.01$          & $9.22_{-0.17}^{+0.18}$        & $6.43_{-0.04}^{+0.03}$  & 0         & $1.43/520$ \\\hline

\multirow{4}{*}{$20^{\dagger}$} & \multirow{4}{*}{826}                                                  & \multirow{4}{*}{$1922$}    & $2.65_{-0.11}^{+0.03}$ & $2.58_{-0.08}^{+0.07}$ & -                      & -                       & $6.76_{-0.07}^{+0.05}$  &   100    & $1.22/503$ \\
     &                                                      &           & $1.82_{-0.03}^{+0.07}$ &	$2$	& $1.08_{-0.06}^{+0.04}$	 & $11.3_{-0.95}^{+1.89}$ &	$6.52_{-0.14}^{+0.06}$& $64\pm11$&	$1.11/502$ \\
     &                                                      &           & $2.14_{-0.03}^{+0.1}$ &	$2.4$ &	$1.12\pm0.07$	& $9.49_{-1.49}^{+1.39}$ &	$6.35_{-0.09}^{+0.07}$&$76\pm12$ &	$1.09/502$   \\
     &                                                      &           & $1.11_{-0.06}^{+0.04}$ & -                      & $1.48_{-0.02}^{+0.07}$ & $8.13_{-0.74}^{+0.57}$  & $5.8_{-0.15}^{+0.05}$   &   0    & $1.75/503$  \\ \hline
\multirow{4}{*}{$21^{\dagger}$} & \multirow{4}{*}{831}                                                  & \multirow{4}{*}{$826$}     & $2.95_{-0.39}^{+0.29}$ & $2.65\pm0.19$          & -                      & -                       & $7.69_{-0.16}^{+0.1}$   &    100   & $1.33/364$   \\
     &                                                      &           & $2.04_{-0.05}^{+0.1}$  & $2$                    & $1.12_{-0.06}^{+0.08}$ & $14.23_{-0.96}^{+2.52}$ & $6.75_{-0.21}^{+0.12}$  & $35\pm10$    & $1.07/363$ \\
     &                                                      &          & $2.26_{-0.15}^{+0.11}$ & $2.4$                  & $1.13_{-0.07}^{+0.09}$ & $13.1_{-2.17}^{+2.36}$  & $6.67_{-0.18}^{+0.16}$  &   $46\pm14$    & $1.06/363$   \\
     &                                                      &           & $1.64\pm0.07$          & -                      & $1.31\pm0.07$          & $12.06_{-1.75}^{+1.32}$ & $6.23_{-0.13}^{+0.07}$  &    0   & $1.21/364$  \\ \hline
\multirow{3}{*}{$22^{\star \dagger}$} & \multirow{3}{*}{850}                                                  & \multirow{3}{*}{$1822$}    &  $2.09_{-0.08}^{+0.07}$ & $2$                    & $0.95\pm0.03$          & $21.91_{-1.86}^{+0.98}$ & $5.84_{-0.08}^{+0.07}$  & $23\pm4$    & $1.14/471$   \\
     &                                                      &           & $2.23_{-0.08}^{+0.04}$ & $2.4$                  & $0.94_{-0.02}^{+0.03}$ & $21.92_{-1.91}^{+2.16}$ & $5.81_{-0.11}^{+0.04}$  &   $32\pm6$    & $1.14/471$  \\
     &                                                      &           & $1.74_{-0.06}^{+0.1}$  & -                      & $1.1_{-0.05}^{+0.03}$  & $17.26_{-1.52}^{+1.58}$ & $5.49_{-0.14}^{+0.04}$  &    0   & $1.45/472$   \\ \hline
\multirow{4}{*}{$23$} & \multirow{3}{*}{860}                                                  & \multirow{3}{*}{$2450$}    &  $2.08\pm0.02$          & $2$           & $1.11\pm0.02$          & $17.17_{-0.42}^{+0.44}$ & $7.76_{-0.05}^{+0.04}$ & $26\pm3$         & $1.22/525$   \\
     &                                                      &           & $2.21\pm0.03$          & $2.4$         & $1.13\pm0.02$          & $15.8_{-0.34}^{+0.35}$  & $7.73\pm0.04$          & $37\pm4$         & $1.22/525$   \\
     &                                                      &           & $1.9\pm0.02$           & -             & $1.23\pm0.01$          & $15.06\pm0.23$          & $7.48\pm0.03$          & 0          & $1.65/526$   \\ \hline

\multirow{3}{*}{$24$} & \multirow{3}{*}{870}                                                  & \multirow{3}{*}{$1326$}      & $2.06\pm0.04$          & $2$           & $1.16\pm0.02$          & $16.16_{-0.50}^{+0.54}$  & $6.99\pm0.06$          & $12\pm5$         & $0.98/449$   \\
     &                                                      &           & $2.12_{-0.06}^{+0.05}$ & $2.4$         & $1.17\pm0.02$          & $15.68_{-0.39}^{+0.42}$ & $6.98\pm0.06$          & $16\pm7$         & $0.98/449$   \\
     &                                                      &           & $1.98\pm0.02$          & -             & $1.21\pm0.01$          & $15.34_{-0.33}^{+0.34}$ & $6.88_{-0.04}^{+0.03}$ & 0          & $1.01/450$   \\ \hline
 \multirow{3}{*}{$25$} & \multirow{3}{*}{880}                                                  & \multirow{3}{*}{$1856$}              & $2.1\pm0.03$           & $2$           & $1.11\pm0.02$          & $18.36_{-0.51}^{+0.54}$ & $7.53_{-0.08}^{+0.04}$ & $14\pm3$         & $0.88/470$   \\
     &                                                      &           & $2.17\pm0.04$          & $2.4$         & $1.12\pm0.02$          & $17.71_{-0.42}^{+0.44}$ & $7.51_{-0.06}^{+0.05}$ & $21\pm5$         & $0.88/470$   \\
     &                                                      &           & $1.99\pm0.02$          & -             & $1.17\pm0.01$          & $16.99_{-0.31}^{+0.32}$ & $7.37_{-0.04}^{+0.03}$ & 0          & $0.98/471$  
\end{tabular}
\end{table*}

\begin{table*}[]
\caption*{Table 1 (Cont.)}
\hspace*{-1cm}\begin{tabular}{ccc|cc|cc|c|c|c}

$ID^a$                  &\begin{tabular}[c]{@{}c@{}}MJD-\\ -56000$^b$\end{tabular} & \begin{tabular}[c]{@{}c@{}}Exp. XRT$^c$\\ s\end{tabular}         & \begin{tabular}[c]{@{}c@{}}$N_H^d$,\\  $10^{22}$cm$^{-2}$\end{tabular}                  & $\Gamma^e$      & \begin{tabular}[c]{@{}c@{}}$T_{in}$, \\ keV $^f$\end{tabular}               &\begin{tabular}[c]{@{}c@{}}$R_{in} cos^{-1/2}(i)$, \\ km $^g$\end{tabular}                   & Flux$^h$               & $f_{po}^j$ & $\chi_N^2/dof$ \\ \hline\hline

\multirow{3}{*}{26} & \multirow{3}{*}{890}                                  & \multirow{3}{*}{2164} & $1.88\pm0.03$          & $2$      & $1.17\pm0.02$ & $16.78_{-0.42}^{+0.45}$ & $7.46_{-0.05}^{+0.06}$ & $3\pm2$          & $1.02/471$   \\
                    &                                                       &                       & $1.89_{-0.04}^{+0.05}$ & $2.4$    & $1.17\pm0.02$ & $16.62_{-0.34}^{+0.35}$ & $7.45_{-0.05}^{+0.07}$ & $4\pm2$          & $1.02/471$   \\
                    &                                                       &                       & $1.85\pm0.02$          & -        & $1.18\pm0.01$ & $16.53\pm0.31$          & $7.43\pm0.05$          & 0          & $1.02/472$   \\ \hline
\multirow{3}{*}{27} & \multirow{3}{*}{900}                                  & \multirow{3}{*}{1909} & $2.05\pm0.03$          & $2$      & $1.12\pm0.02$ & $16.69_{-0.40}^{+0.41}$ & $6.54_{-0.04}^{+0.03}$ & $12\pm3$         & $1.07/499$   \\
                    &                                                       &                       & $2.1\pm0.04$           & $2.4$    & $1.13\pm0.01$ & $16.20_{-0.33}^{+0.34}$ & $6.53_{-0.04}^{+0.03}$ & $17\pm4$         & $1.07/499$   \\
                    &                                                       &                       & $1.96\pm0.02$          & -        & $1.18\pm0.01$ & $15.67_{-0.26}^{+0.27}$ & $6.44_{-0.03}^{+0.04}$ & 0          & $1.15/500$   \\ \hline
\multirow{3}{*}{$28^{\star}$} & \multirow{3}{*}{910}                                  & \multirow{3}{*}{982}  &  $1.98\pm0.04$          & $2$           & $1.1\pm0.02$           & $17.46_{-0.58}^{+0.62}$       & $6.24_{-0.05}^{+0.06}$  & $9\pm4$   & $1.12/417$ \\
&& & $2.02\pm0.05$          & $2.4$         & $1.1\pm0.02$           & $17.08_{-0.49}^{+0.52}$       & $6.23_{-0.07}^{+0.06}$  & $13\pm6$  & $1.12/417$ \\
&&& $1.91\pm0.03$          & -             & $1.14\pm0.01$          & $16.56_{-0.38}^{+0.39}$       & $6.15_{-0.06}^{+0.04}$  & 0         & $1.15/418$ \\\hline
\multirow{3}{*}{29} & \multirow{3}{*}{920}                                  & \multirow{3}{*}{638}  & $1.98\pm0.05$          & $2$      & $1.05\pm0.03$ & $17.87_{-0.79}^{+0.86}$ & $5.21_{-0.07}^{+0.05}$ & $7\pm5$          & $1.0/342$    \\
                    &                                                       &                       & $2.02\pm0.07$          & $2.4$    & $1.06\pm0.03$ & $17.59_{-0.67}^{+0.73}$ & $5.21_{-0.07}^{+0.06}$ & $10\pm6$         & $1.0/342$    \\
                    &                                                       &                       & $1.93\pm0.03$          & -        & $1.09\pm0.01$ & $17.08_{-0.51}^{+0.53}$ & $5.15\pm0.04$          & 0          & $1.01/343$   \\ \hline
\multirow{3}{*}{$30^{\dagger}$} & \multirow{3}{*}{924}                                  & \multirow{3}{*}{857}  & $2.23_{-0.15}^{+0.13}$          & $2$      & $0.94_{-0.05}^{+0.06}$ & $22.90_{-3.34}^{+3.17}$ & $4.69_{-0.18}^{+0.00}$ & $6\pm3$         & $1.08/373$   \\
                    &                                                       &                       & $2.25_{-0.13}^{+0.15}$          & $2.4$    & $0.95_{-0.03}^{+0.05}$ & $22.19_{-2.54}^{+3.70}$ & $4.68_{-0.30}^{+0.05}$ & $10\pm5$         & $1.07/373$   \\
                    &                                                       &                       & $2.11_{-0.05}^{+0.13}$          & -        & $0.99_{-0.05}^{+0.04}$  & $20.82_{-2.28}^{+3.25}$ & $4.62_{-0.09}^{+0.02}$ & 0          & $1.09/374$   \\ \hline
\multirow{3}{*}{$31^{\star}$} & \multirow{3}{*}{932}                                  & \multirow{3}{*}{2073} & $1.96\pm0.03$          & $2$           & $1.0\pm0.02$           & $17.13_{-0.53}^{+0.56}$       & $3.73\pm0.03$           & $8\pm3$   & $1.15/424$ \\
&&& $2.0\pm0.04$           & $2.4$         & $1.0\pm0.02$           & $16.92_{-0.48}^{+0.51}$       & $3.73_{-0.03}^{+0.02}$  &           & $1.15/424$ \\
&&& $1.9\pm0.02$           & -             & $1.04\pm0.01$          & $16.12_{-0.34}^{+0.35}$       & $3.68_{-0.03}^{+0.02}$  & 0         & $1.2/425$ \\ \hline

\multirow{3}{*}{32} & \multirow{3}{*}{940}                                  & \multirow{3}{*}{1680} & $2.13\pm0.03$          & $2$      & $0.96\pm0.01$ & $19.70_{-0.61}^{+0.65}$ & $3.90_{-0.04}^{+0.02}$ & $9\pm2$          & $1.04/415$   \\
                    &                                                       &                       & $2.18\pm0.04$          & $2.4$    & $0.95\pm0.01$ & $19.48_{-0.58}^{+0.62}$ & $3.89\pm0.03$          & $15\pm4$         & $1.03/415$   \\
                    &                                                       &                       & $2.05\pm0.02$          & -        & $1.01\pm0.01$ & $18.04_{-0.39}^{+0.40}$ & $3.82_{-0.03}^{+0.02}$ & 0          & $1.14/416$   \\ \hline
\multirow{3}{*}{35} & \multirow{3}{*}{953}                                  & \multirow{3}{*}{1793} & $2.06\pm0.03$          & $2$      & $0.92\pm0.02$ & $19.03_{-0.67}^{+0.70}$ & $3.27_{-0.03}^{+0.02}$ & $14\pm2$         & $1.08/404$   \\
                    &                                                       &                       & $2.12\pm0.04$          & $2.4$    & $0.92\pm0.02$ & $18.60_{-0.62}^{+0.66}$ & $3.26\pm0.03$          & $21\pm4$         & $1.08/404$   \\
                    &                                                       &                       & $1.94\pm0.02$          & -        & $0.99\pm0.01$ & $16.68_{-0.37}^{+0.39}$ & $3.17\pm0.02$          & 0          & $1.27/405$  \\\hline
\multirow{3}{*}{36} & \multirow{3}{*}{954}                                 & \multirow{3}{*}{1751} & $1.92\pm0.03$          & $2$      & $0.93\pm0.02$ & $18.66_{-0.63}^{+0.66}$ & $3.36_{-0.03}^{+0.02}$ & $11\pm3$         & $1.12/400$   \\
                    &                                                      &                       & $1.96\pm0.04$          & $2.4$    & $0.94\pm0.02$ & $18.28_{-0.57}^{+0.61}$ & $3.35_{-0.03}^{+0.02}$ & $17\pm4$         & $1.12/400$   \\
                    &                                                      &                       & $1.83\pm0.02$          & -        & $0.99\pm0.01$ & $16.89_{-0.37}^{+0.38}$ & $3.28\pm0.02$          & 0          & $1.24/401$   \\ \hline
\multirow{3}{*}{37} & \multirow{3}{*}{955}                                 & \multirow{3}{*}{1931} & $2.02\pm0.03$          & $2$      & $0.9\pm0.02$  & $19.51_{-0.68}^{+0.71}$ & $3.14_{-0.03}^{+0.02}$ & $15\pm2$         & $1.03/409$   \\
                    &                                                      &                       & $2.08\pm0.03$          & $2.4$    & $0.9\pm0.02$  & $19.05_{-0.64}^{+0.68}$ & $3.13_{-0.03}^{+0.02}$ & $24\pm4$         & $1.03/409$   \\
                    &                                                      &                       & $1.89\pm0.02$          & -        & $0.98\pm0.01$ & $16.8_{-0.37}^{+0.38}$  & $3.03\pm0.02$          & 0          & $1.27/410$   \\ \hline
\multirow{3}{*}{38} & \multirow{3}{*}{956}                                 & \multirow{3}{*}{1920} & $2.09\pm0.03$          & $2$      & $0.88\pm0.01$ & $20.73_{-0.73}^{+0.77}$ & $3.28_{-0.02}^{+0.03}$ & $18\pm2$         & $1.08/409$   \\
                    &                                                      &                       & $2.16\pm0.03$          & $2.4$    & $0.88\pm0.02$ & $20.26_{-0.71}^{+0.76}$ & $3.27_{-0.03}^{+0.02}$ & $26\pm3$         & $1.07/409$   \\
                    &                                                      &                       & $1.94\pm0.02$          & -        & $0.98\pm0.01$ & $17.15_{-0.38}^{+0.39}$ & $3.15\pm0.02$          & 0          & $1.48/410$   \\ \hline
\multirow{3}{*}{39} & \multirow{3}{*}{959.8}                               & \multirow{3}{*}{540}  & $2.09\pm0.07$          & $2$      & $0.89\pm0.03$ & $19.23_{-1.43}^{+1.61}$ & $2.86_{-0.07}^{+0.04}$ & $16\pm5$         & $0.96/208$   \\
                    &                                                      &                       & $2.15\pm0.08$          & $2.4$    & $0.89\pm0.03$ & $18.79_{-1.34}^{+1.54}$ & $2.85_{-0.07}^{+0.04}$ & $24\pm8$         & $0.95/208$   \\
                    &                                                      &                       & $1.95\pm0.05$          & -        & $0.97\pm0.02$ & $16.35_{-0.73}^{+0.77}$ & $2.75\pm0.03$          & 0          & $1.06/209$   \\ \hline
\multirow{3}{*}{40} & \multirow{3}{*}{960.3}                               & \multirow{3}{*}{2055} & $1.91\pm0.03$          & $2$      & $0.89\pm0.01$ & $18.36_{-0.64}^{+0.67}$ & $2.62\pm0.02$          & $13\pm2$         & $1.0/387$    \\
                    &                                                      &                       & $1.96\pm0.03$          & $2.4$    & $0.89\pm0.02$ & $18.01_{-0.61}^{+0.65}$ & $2.62_{-0.03}^{+0.02}$ & $20\pm3$         & $1.01/387$   \\
                    &                                                      &                       & $1.80\pm0.02$          & -        & $0.96\pm0.01$ & $16.03_{-0.37}^{+0.38}$ & $2.54_{-0.02}^{+0.01}$ & 0          & $1.23/388$   \\ \hline
\multirow{3}{*}{41} & \multirow{3}{*}{961}                                 & \multirow{3}{*}{1018} & $1.73\pm0.04$          & $2$      & $0.94\pm0.02$ & $16.36_{-0.77}^{+0.83}$ & $2.60_{-0.04}^{+0.02}$ & $7\pm4$          & $1.09/314$   \\
                    &                                                      &                       & $1.75_{-0.06}^{+0.05}$ & $2.4$    & $0.94\pm0.02$ & $16.1_{-0.69}^{+0.75}$  & $2.59_{-0.04}^{+0.03}$ & $10\pm6$         & $1.09/314$   \\
                    &                                                      &                       & $1.68\pm0.03$          & -        & $0.97\pm0.01$ & $15.41_{-0.49}^{+0.50}$ & $2.56\pm0.02$          & 0          & $1.11/315$   \\ \hline
\multirow{3}{*}{42} & \multirow{3}{*}{962}                                 & \multirow{3}{*}{809}  & $1.82\pm0.05$          & $2$      & $0.89\pm0.03$ & $17.91_{-1.06}^{+1.18}$ & $2.37_{-0.05}^{+0.03}$ & $8\pm5$          & $1.04/259$   \\
                    &                                                      &                       & $1.84_{-0.07}^{+0.06}$ & $2.4$    & $0.89\pm0.03$ & $17.6_{-0.96}^{+1.07}$  & $2.37_{-0.04}^{+0.03}$ & $12\pm8$         & $1.05/259$   \\
                    &                                                      &                       & $1.75\pm0.04$          & -        & $0.93\pm0.01$ & $16.58_{-0.62}^{+0.65}$ & $2.32_{-0.03}^{+0.02}$ & 0          & $1.07/260$  
\end{tabular}
\caption*{a - XRT observation number, b - observation time, MJD 56000; c - XRT exposure; d - interstellar absorption; e-photon index; f-inner disk temperature; g-inner disk radius for a distance to the system of 8.5 kpc; h-total absorbed 0.8-10 keV flux of the model in units of $10^{-9}$ erg cm$^{-2}$ s$^{-1}$; i-contribution of the power-law component to the total flux 0.8-10 keV flux. The symbol $\dagger$ marks the
spectra in fitting which the gain fit command was used (the slope and offset are given in Table 2). The symbol $\star$ marks the observations
in which the position-sensitive response matrix was used (see the text).}

\end{table*}

\begin{table}
\caption{Parameters of the gain fit command used in fitting
the XRT spectra}
\label{tbl-gain-pars}
\begin{tabular}{l|l|ll}
ID                    & Model                 & Slope                  & Offset, keV             \\ \hline
\multirow{4}{*}{$4$}  & po                     & $1.02\pm0.01$          & $0.03\pm0.02$           \\
                      & $(po+disk) \Gamma=1.5$ & $0.97_{-0.0}^{+0.01}$  & $0.1\pm0.02$            \\
                      & $(po+disk) \Gamma=1.8$ & $1.0\pm0.01$           & $0.05\pm0.02$           \\
                      & $(po+disk) \Gamma=2$   & $1.02_{-0.0}^{+0.01}$  & $0.02\pm0.02$           \\ \hline
\multirow{4}{*}{$5$}  & po                     & $1.02\pm0.01$          & $0.06_{-0.03}^{+0.02}$  \\
                      & $(po+disk) \Gamma=1.5$ & $0.99\pm0.01$          & $0.12\pm0.02$           \\
                      & $(po+disk) \Gamma=1.8$ & $1.02_{-0.01}^{+0.0}$  & $0.05_{-0.01}^{+0.02}$  \\
                      & $(po+disk) \Gamma=2$   & $1.04_{-0.0}^{+0.01}$  & $0.01\pm0.02$           \\ \hline
\multirow{4}{*}{$9$}  & po                     & $1.03\pm0.02$          & $0.04_{-0.03}^{+0.04}$  \\
                      & $(po+disk) \Gamma=1.5$ & $0.99\pm0.01$          & $0.12_{-0.02}^{+0.03}$  \\
                      & $(po+disk) \Gamma=1.8$ & $1.0\pm0.01$           & $0.08_{-0.03}^{+0.02}$  \\
                      & $(po+disk) \Gamma=2$   & $1.02\pm0.01$          & $0.04\pm0.02$           \\ \hline
\multirow{4}{*}{$11$} & po                     & $0.89\pm0.01$          & $0.18_{-0.03}^{+0.02}$  \\
                      & $(po+disk) \Gamma=2$   & $1.04_{-0.01}^{+0.02}$ & $-0.01\pm0.03$          \\
                      & $(po+disk) \Gamma=2.4$ & $1.05_{-0.02}^{+0.01}$ & $-0.03_{-0.03}^{+0.04}$ \\
                      & disk                   & $1.04_{-0.01}^{+0.02}$ & $-0.0_{-0.04}^{+0.03}$  \\ \hline
\multirow{4}{*}{$14$} & po                     & $1.0_{-0.01}^{+0.02}$  & $-0.04_{-0.02}^{+0.03}$ \\
                      & $(po+disk) \Gamma=2$   & $1.07_{-0.01}^{+0.02}$ & $-0.06\pm0.02$          \\
                      & $(po+disk) \Gamma=2.4$ & $1.08_{-0.01}^{+0.02}$ & $-0.09_{-0.03}^{+0.02}$ \\
                      & disk                   & $1.08_{-0.02}^{+0.01}$ & $-0.03_{-0.02}^{+0.03}$ \\ \hline
\multirow{4}{*}{$17$} & po                     & $0.95\pm0.01$          & $-0.0_{-0.0}^{+0.04}$   \\
                      & $(po+disk) \Gamma=2$   & $1.04_{-0.02}^{+0.01}$ & $-0.05_{-0.02}^{+0.04}$ \\
                      & $(po+disk) \Gamma=2.4$ & $1.04\pm0.01$          & $-0.07_{-0.02}^{+0.03}$ \\
                      & disk                   & $1.03_{-0.02}^{+0.01}$ & $-0.01\pm0.03$          \\ \hline
\multirow{4}{*}{$20$} & po                     & $1.04_{-0.02}^{+0.01}$ & $-0.08\pm0.03$          \\
                      & $(po+disk) \Gamma=2$   & $1.02\pm0.01$          & $0.02_{-0.03}^{+0.02}$  \\
                      & $(po+disk) \Gamma=2.4$ & $1.05\pm0.01$          & $-0.04_{-0.02}^{+0.01}$ \\
                      & disk                   & $1.03_{-0.02}^{+0.01}$ & $0.1_{-0.03}^{+0.04}$   \\ \hline
\multirow{4}{*}{$21$} & po                     & $1.01_{-0.01}^{+0.02}$ & $-0.03\pm0.06$          \\
                      & $(po+disk) \Gamma=2$   & $1.07_{-0.03}^{+0.02}$ & $-0.06_{-0.04}^{+0.05}$ \\
                      & $(po+disk) \Gamma=2.4$ & $1.08\pm0.02$          & $-0.08\pm0.04$          \\
                      & disk                   & $1.08_{-0.03}^{+0.01}$ & $-0.03_{-0.02}^{+0.06}$ \\ \hline
\multirow{3}{*}{$22$}                       & $(po+disk) \Gamma=2$   & $1.06\pm0.02$          & $-0.1\pm0.03$           \\
                      & $(po+disk) \Gamma=2.4$ & $1.07\pm0.02$          & $-0.12_{-0.02}^{+0.03}$ \\
                      & disk                   & $1.07\pm0.02$          & $-0.07\pm0.03$          \\ \hline
\multirow{3}{*}{$30$} & $(po+disk) \Gamma=2$   & $1.08\pm0.03$          & $-0.03\pm0.01$          \\
                      & $(po+disk) \Gamma=2.4$ & $1.07_{-0.01}^{+0.03}$ & $-0.03\pm0.01$          \\
                      & disk                   & $1.07_{-0.02}^{+0.03}$ & $-0.02\pm0.1$          
\end{tabular}
\end{table}

When fitting the broadband spectra, we added the
cross-calibration constants between the three instruments
to take into account the nonsimultaneity of the
observations. A difference of the cross-calibration
constants between the JEM-X and ISGRI/IBIS instruments
is observed when working with the INTEGRAL
data (see, e.g., Filippova et al. 2014a).

\begin{table*}[]
\centering
\caption{Times and exposures of observations for the
broadband spectra obtained from INTEGRAL and XRT
data}
\label{tbl-broadband-pars}
\begin{tabular}{cc|ccc}
$ID^a$ & MJD-56000$^b$ & \begin{tabular}[c]{@{}c@{}}Exp. \\ $XRT$\\ s\end{tabular} & \begin{tabular}[c]{@{}c@{}}Exp. \\$ ISGRI$\\ s\end{tabular} & \begin{tabular}[c]{@{}c@{}}Exp.\\ $JEM-X$\\ s\end{tabular} \\ \hline
$01$  & 736                                                   & $79$                                                      & 4735                                                        & 11244                                                       \\ \hline
$03$  & 742                                                   & $1925$                                                    & 44637                                                       & 10080                                                       \\ \hline
$05$  & 751                                                   & $1832$                                                    & 4219                                                        & 5006                                                        \\ \hline
$07$  & 761                                                   & $1613$                                                    & 25806                                                       & 68580                                                       \\ \hline
$09$  & 771                                                   & $1872$                                                    & 5791                                                        & 10738                                                      
\end{tabular}
\caption*{a-XRT observation number; b-time of observation,
MJD 56000.}
\end{table*}

\section{DATA ANALYSIS}

\subsection{Light Curve during the 2014 Outburst}

\begin{figure*}[t]
\centering
\includegraphics[width=\textwidth]{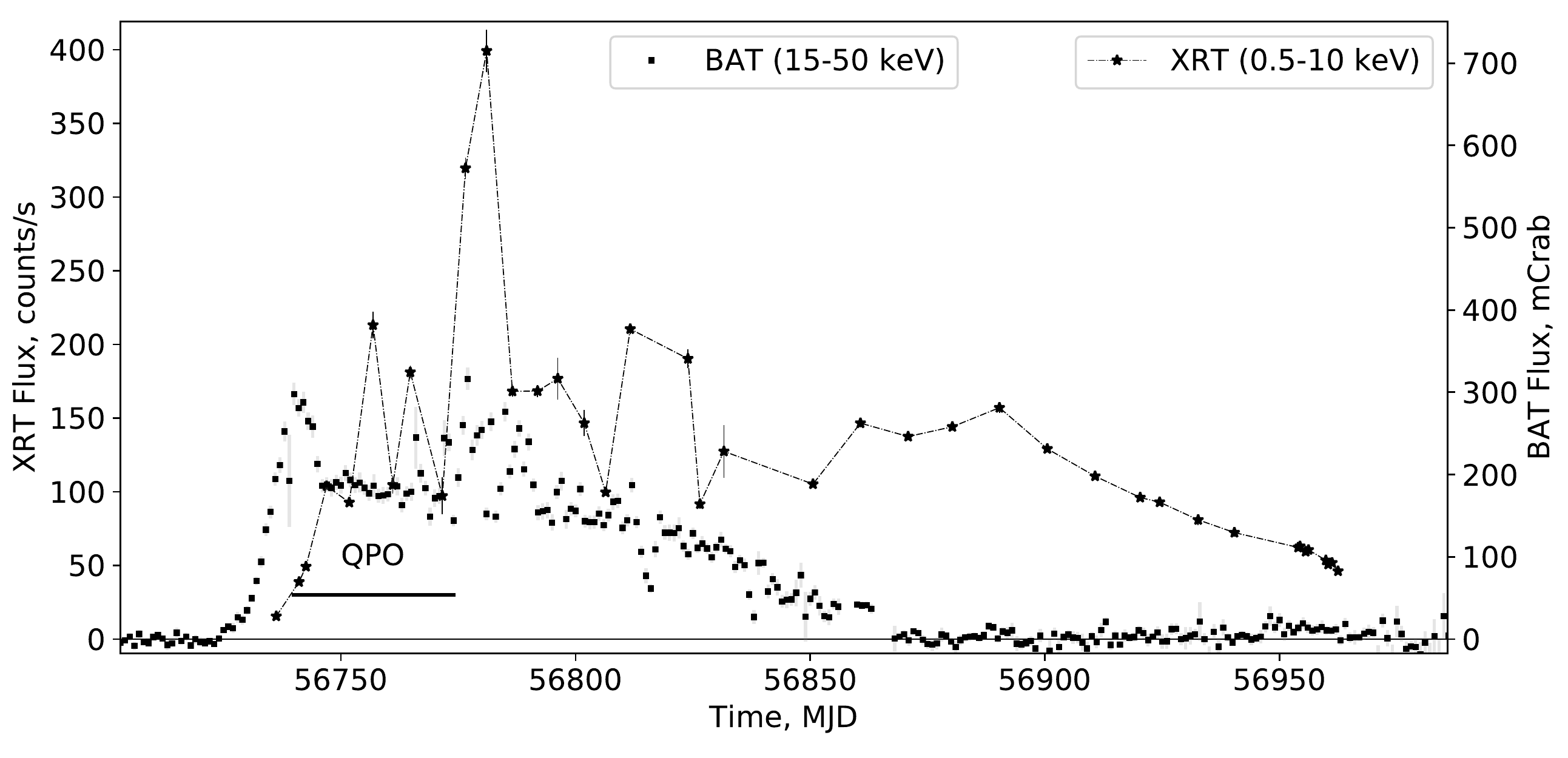}
\caption{Light curve of GRS 1739-278 during the 2014 outburst from Swift/XRT (designated by stars, the data were averaged over one observation) and Swift/BAT (designated by black squares, the data were averaged over one day) data. The horizontal line indicates the period of QPO observations in the variability power spectra for the source (observations 02-09).}
\label{fig-lc1}
\end{figure*}

Figure 1 shows the source's light curve from
Swift/XRT and Swift/BAT data in the 0.5-10 (the
data points were averaged over one observation) and
15-50 keV (the data points were averaged over one
day) energy bands, respectively. To convert the 15-50 keV flux to units corresponding to the flux from
the Crab Nebula, we used the relation 1 Crab =0.22 counts s$^{-1}$ cm$^{-2}$. It can be seen from the figure
that in the 15-50 keV energy band the flux from the
source reached a maximum $\sim0.3$ Crab $\sim15$ days after
the outburst onset, it then dropped by a factor of 1.5
in 5 days and was approximately at a constant level
for $\sim25$ days, after which it began to exhibit a peak-shaped
variability that lasted for about 50 days and passed into the outburst decay phase. The emission
from the source in this energy band ceased to be
recorded $\sim140$ days later.

In the 0.5-10 keV energy band the outburst
reached its maximum ($\sim$1.1 Crab) with a delay relative
to the maximum in the 15-50 keV energy band,
55 days after the onset, but the source exhibited a
peak-shaped flux variability almost immediately from
the outburst onset. On completion of this activity,
135 days after the outburst onset, the flux reached a
constant level $\sim$150 counts s$^{-1}$ ($\sim$400 mCrab) and
remained so for $\sim$30 days, after which it began to
drop. The observations ceased on $\sim$240 day of the
outburst, because the source fell into a region near
the Sun inaccessible to observations. At this time the
0.5-10 keV flux was $\sim$50 counts s$^{-1}$ ($\sim$140 mCrab).
In Fig. 2a the flux in the soft energy band (0.5-
10 keV) is plotted against the source's hardness (the
ratio of the 4-10 and 0.5-4 keV fluxes). It can be
seen from the figure that the diagram has a shape that
resembles the upper part of a typical "q"$\,$ diagram.

\subsection{Spectral Analysis during the 2014 Outburst}

When fitting the spectra, we used typical models
describing the source's spectrum: (1) in the
low/hard state-a power law with a high-energy
cutoff and low-energy absorption, phabs*cutoffpl
(or in the case of fitting only the Swift/XRT data-
a power law and low-energy absorption, phabs*powerlaw); (2) in the intermediate states-a power
law with a high-energy cutoff, a multitemperature
accretion disk, and low-energy absorption, phabs*(cutoffpl + diskbb) (or in the case of fitting only the
Swift/XRT data-a power law, a multitemperature accretion disk, and low-energy absorption, phabs*(powerlaw + diskbb)); (3) in the high/soft state-
a multitemperature accretion disk and low-energy
absorption, phabs*diskbb, or the previous model,
phabs*(powerlaw + diskbb). The quality of the
available data does not allow us to apply the more
complex spectral model including the reflection of
Comptonized radiation from the accretion disk that
was used in fitting the NuSTAR data (Miller et al.
2015; Mereminskiy et al. 2019).

The results of fitting the Swift/XRT spectra are
presented in Table 1. The errors in the parameters
are given for a 90\% confidence interval. The table
also provides the contribution of the unabsorbed
power-law component to the total unabsorbed flux,
the absorbed 0.8-10 keV flux, and the inner accretion
disk radius estimated from the normalization in the
diskbb model $N = (R_{in}/D_{10 kpc})^2 cos i$, where $R_{in}$ is
the "apparent"$\,$ inner disk radius in km, $D_{10 kpc}$ is the
distance to the source in units of 10 kpc, and $i$ is the
inclination to the plane of the sky. The distance to the
source was taken to be 8.5 kpc.

When fitting the spectra obtained only from the
Swift/XRT data, the available 0.8-10 keV energy
band does not allow unambiguous constraints to be
placed on the parameters of the power-law component
in the case of using the multicomponent phabs*(powerlaw + diskbb) model. Therefore, when fitting
the data at the initial outburst phases (from observation
03 to 09), we fixed the photon index at 1.5 and
1.8, which were derived when fitting the broadband
spectra, at 2, and, in the subsequent observations, at
2 and 2.4, typical for the intermediate state (Remillard
and McClintock 2006; Belloni and Motta 2016).

\begin{figure*}[t]

\centering
\includegraphics[width=0.75\textwidth]{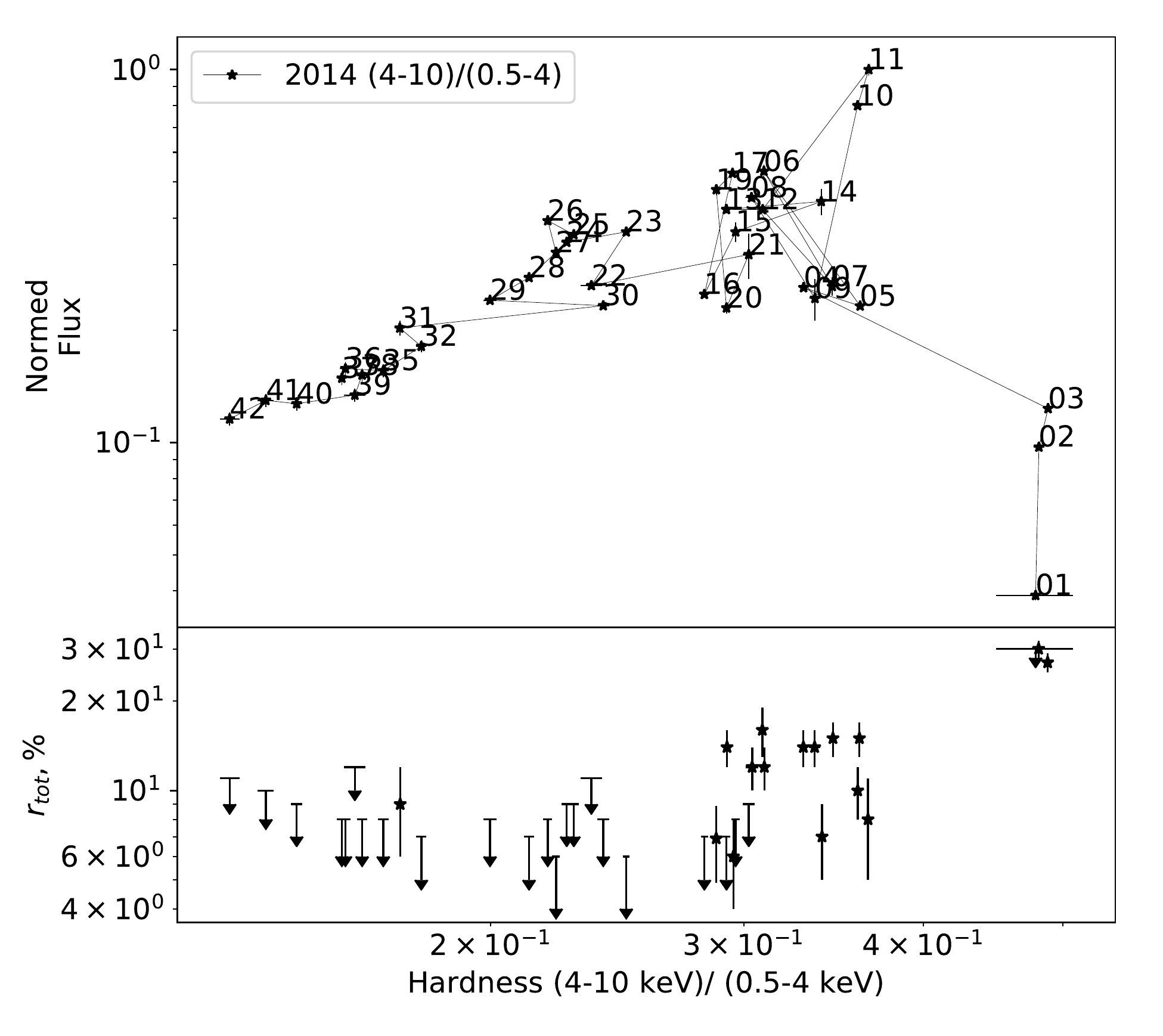}
\caption{Hardness-intensity and hardness-rms diagrams for the 2014 outburst. The hardness defined as the ratio of the count rates in the 4-10 and 0.5-4 keV energy bands is along the horizontal axis. (a) Swift/XRT photon count rate (in the 0.5-10 keV energy band) normalized to the maximum versus hardness. (b) Total fractional rms in the 0.5-10 keV energy band and
the frequency range 0.01-50 Hz versus hardness. Each data point on the upper panel is labeled according to the observation
number.}
\label{fig-hid1}
\end{figure*}

To fit the broadband spectra, we also used the
phabs*highecut*simpl*diskbb model (the parameter
$E_c$ in the highecut model was frozen at a
minimum value of 0.0001 keV, which allowed the
cutoffpl model to be imitated) that takes into account
the physical cutoff of the power-law component
at low energies in a simplified way. When estimating
the errors in the parameters of the phabs*highecut*simpl*diskbb model, we fixed the absorption $N_H$
(and the parameter $f_{scat}$ for observation 03) at their
values found. The results of fitting the broadband
spectra are presented in Table 4. It can be seen
from the table that this model gives a systematically
larger inner accretion disk radius than does the model
with the cutoffpl component, while the remaining
parameters do not differ greatly.

It can be seen from Tables 1 and 4 that in observations
01-03 the model with a power law describes
well the spectra, while in observation 03 the system's spectrum can also be described by the power-law
model with a multitemperature disk (the $\chi^2$ value is
almost the same for the phabs*(powerlaw +diskbb)
and phabs*powerlaw models). Beginning from the
fourth observation, fitting the data by the model of
a multitemperature disk with a power law is more
preferable than that by the model only with a power
law. Up to observation 23, fitting the data by a power
law or a multitemperature disk with low-energy absorption
gives a $\chi^2$ value systematically poorer than
does the multicomponent model. In this case, the
low-energy absorption, the temperature, and the inner
radius of the accretion disk depend on the chosen
photon index, i.e., the available data allow only the
range of values (given in Table 1) in which the model
parameters can lie to be specified.

\begin{table*}[t]
\caption{Parameters of the best-fit models $phabs*powerlaw$, $phabs*(diskbb+cutoffpl)$ and $phabs*highecut*simpl*diskbb$ with fixed $E_c=0.0001$ keV for the broadband energy spectra of GRS 1739-278.}
\label{tbl-broadband}
\resizebox{\textwidth}{!}{%
\hspace*{-1cm}\begin{tabular}{c|cc|cccc|cc|cc|c}
$ID$                  & $C_{JEM-X}^a$          & $C_{XRT}^a$            & \begin{tabular}[c]{@{}c@{}}$N_H$, \\ $10^{22}$cm${^2}^b$\end{tabular} & $\Gamma^{ c}$          & \begin{tabular}[c]{@{}c@{}}$E_{cut}$, \\ keV $ ^d$\end{tabular} & $f_{scat}^e$  & \begin{tabular}[c]{@{}c@{}}$T_{in}$,\\ keV $^f$\end{tabular} &  \begin{tabular}[c]{@{}c@{}}$R_{in} cos^{-1/2}(i)$,\\ km $^g$\end{tabular} & \begin{tabular}[c]{@{}c@{}}$f_{po}$, \\ $\%^h$\end{tabular} & Flux$^j$           & $\chi^2/dof$ \\ \hline
$01$                  & $0.51_{-0.08}^{+0.09}$ & $1.24_{-0.34}^{+0.46}$ & $1.9_{-0.42}^{+0.46}$                                                 & $1.45_{-0.21}^{+0.22}$ & $95_{-29}^{+74}$                                                &               & -                                                            & -                               & 100                                                         & $1.2_{-0.5}^{+0.2}$ & $1.76/17$    \\ \hline
\multirow{3}{*}{$03$} & $0.74\pm0.05$          & $1.42_{-0.11}^{+0.12}$ & $1.6\pm0.04$                                                          & $1.2\pm0.03$           & $20.7_{-1.2}^{+1.4}$                                            &               & -                                                            & -                               & 100                                                         & $3.1\pm0.2$         & $1.22/506$   \\
                      & $0.74\pm0.05$          & $1.42_{-0.12}^{+0.13}$ & $1.84_{-0.15}^{+0.17}$                                                & $1.21\pm0.05$          & $20.6_{-1.4}^{+1.5}$                                            &               & $0.33_{-0.05}^{+0.09}$                                       & $46_{-28}^{+48}$                & 93                                                          & $3.1_{-0.3}^{+0.2}$ & $1.21/504$   \\
                 & $0.73\pm0.05$          & $1.37_{-0.11}^{+0.12}$ & $2.05$                                                                & $1.26_{-0.02}^{+0.03}$ & $21.4_{-1.2}^{+1.3}$                                            & 0.68          & $0.28\pm0.01$                                                & $211_{-22}^{+25}$               &                                                             & $3.2_{-0.3}^{+0.2}$ & $1.22/506$   \\ \hline
\multirow{3}{*}{$05$} & $0.81_{-0.07}^{+0.08}$ & $1.8_{-0.16}^{+0.18}$  & $1.83\pm0.03$                                                         & $1.8\pm0.03$           & $35.9_{-3.9}^{+4.7}$                                            &               & -                                                            & -                               & 100                                                         & $3.9\pm0.3$         & $1.64/564$   \\
                      & $0.84_{-0.07}^{+0.08}$ & $2.19_{-0.21}^{+0.24}$ & $2.04\pm0.06$                                                         & $1.51_{-0.07}^{+0.06}$ & $25.0_{-2.5}^{+2.9}$                                            &               & $0.47_{-0.03}^{+0.04}$                                       & $38.8_{-7.5}^{+10.0}$           & 80                                                          & $3.3_{-0.5}^{+0.2}$ & $1.33/562$   \\
              & $0.84_{-0.07}^{+0.08}$ & $2.21_{-0.21}^{+0.24}$ & $1.93$                                                                & $1.5\pm0.06$           & $24.9_{-2.4}^{+2.9}$                                            & $0.58\pm0.01$ & $0.47\pm0.02$                                                & $64.5_{-6.8}^{+8.2}$            &                                                             & $3.3_{-0.3}^{+0.2}$ & $1.33/563$   \\ \hline
\multirow{3}{*}{$07$}   & $0.64\pm0.03$          & $1.54_{-0.09}^{+0.1}$  & $2.15\pm0.03$                                                         & $2.03\pm0.03$          & $38.5_{-2.8}^{+3.1}$                                            &               & -                                                            & -                               & 100                                                         & $5.3_{-0.3}^{+0.4}$ & $1.61/524$   \\
                      & $0.67_{-0.03}^{+0.04}$ & $1.79_{-0.12}^{+0.13}$ & $2.43_{-0.08}^{+0.09}$                                                & $1.88\pm0.05$          & $32.2_{-2.4}^{+2.7}$                                            &               & $0.38\pm0.03$                                                & $84.5_{-21.2}^{+29.5}$          & 81                                                          & $4.7_{-0.4}^{+0.3}$ & $1.43/522$   \\                & $0.67_{-0.03}^{+0.04}$ & $1.8_{-0.12}^{+0.13}$  & $2.34$                                                                & $1.87\pm0.05$          & $31.9_{-2.3}^{+2.6}$                                            & $0.5\pm0.02$  & $0.38\pm0.02$                                                & $137.1_{-16.6}^{+20.6}$         &                                                             & $4.6_{-0.4}^{+0.2}$ & $1.43/523$   \\ \hline
\multirow{3}{*}{$09$}               & $0.65\pm0.05$          & $1.22_{-0.10}^{+0.11}$ & $1.95\pm0.03$                                                         & $1.97\pm0.03$          & $37.0_{-3.7}^{+4.4}$                                            &               & -                                                            & -                               & 100                                                         & $5.3\pm0.4$         & $1.32/516$   \\
                      & $0.71_{-0.05}^{+0.06}$ & $1.54_{-0.14}^{+0.16}$ & $2.08_{-0.06}^{+0.07}$                                                & $1.67\pm0.07$          & $26.5_{-2.6}^{+3.0}$                                            &               & $0.49\pm0.04$                                                & $40.4_{-8.1}^{+11.0}$           & 82                                                          & $4.3_{-0.6}^{+0.3}$ & $1.11/514$   \\             & $0.71_{-0.05}^{+0.06}$ & $1.54_{-0.14}^{+0.15}$ & $1.97$                                                                & $1.67_{-0.07}^{+0.06}$ & $26.5_{-2.5}^{+2.9}$                                            & $0.55\pm0.02$ & $0.48\pm0.03$                                                & $71.2_{-8.4}^{+10.3}$           &                                                             & $4.2_{-0.5}^{+0.3}$ & $1.1/515$   
\end{tabular}}

{a-cross-calibration constants for JEM-X and XRT relative to ISGRI, respectively; b-interstellar absorption; c-photon index; d-
cutoff energy of the cutoffpl model; e-disk fraction subject to Comptonization (simpl); f-accretion disk temperature; g-inner disk
radius for a distance to the system of 8.5 kpc; h-contribution of the power-law component to the total 0.8-10 keV flux; i-absorbed
0.8-10 keV flux of the broadband model, in units of $10^{-9}$ erg cm$^{-2}$ s$^{-1}$. The errors are given for a 90\% confidence interval.}
\end{table*}

In observations 24-42 the data are well fitted
by the model of a multitemperature disk with low-energy
absorption. Although adding the power-law component when fitting some of the spectra formally
reduces the $\chi^2$ value, the contribution of the powerlaw
component is so small ($≤30\%$) that a change in
the photon index within the range 2-2.4 has virtually
no effect on the remaining model parameters.

\subsection{Temporal Variability during the 2014 Outburst}

To analyze the source's variability, we constructed
its power spectra in several energy bands: 0.5-10 (F),
0.5-3 (A), and 3-10 keV (B).
QPOs were detected in the power spectra constructed
for observations 02-09. As the best-fit
model for the power spectra obtained in these observations
we used a model consisting of two Lorentz
profiles (one Lorentz profile described the broadband To analyze the source's variability, we constructed
its power spectra in several energy bands: 0.5-10 (F),
0.5-3 (A), and 3-10 keV (B).
QPOs were detected in the power spectra constructed
for observations 02-09. As the best-fit
model for the power spectra obtained in these observations
we used a model consisting of two Lorentz
profiles (one Lorentz profile described the broadband

\begin{equation*}
\begin{array}{l}
P(f)=\frac{N_{qpo}}{\pi} \frac{\delta f_{qpo}/2}{(f-f_{qpo})^2+(\delta f_{qpo}/2)^2}+\\\frac{N_{sub}}{\pi} \frac{\delta f_{sub}/2}{(f-f_{sub})^2+(\delta f_{sub}/2)^2}+P_{noise}
\end{array}
\end{equation*}

where $f_{qpo}$ and $\delta f_{qpo}$  are the frequency and width of
the Lorentzian responsible for the QPOs, $f_{sub}$ and $\delta f_{sub}$ are the frequency and width of the Lorentzian
responsible for the broadband noise, $N_{qpo}$ and $N_{sub}$
are the normalizations of the QPO and broadband
noise components, $P_{noise}$ is the constant responsible
for the Poisson noise level. In the subsequent analysis
we assumed that $f_{sub}$ = 0. This model describes well
the power spectra of black hole candidates in the  low/hard and low intermediate states (Belloni and
Motta 2016).

To determine the parameters of the best-fit model,
we used the maximum likelihood method (see Leahy
et al. 1983; Vikhlinin et al. 1994). As the likelihood
function we used the product of the probability density
functions for a $\chi^2$ distribution with 2n degrees of
freedom:

\begin{equation}
L=\prod{f_{\chi^2_{2n}}({{ P_{i,src} 2n}\over{P_{i,model}}})},
\end{equation}
where $P_{i,src}$ is the measured rms power of the source
in the i-th frequency bin, $P_{i,model}$ is the rms power of
the source in the same bin obtained from the model,
and n is the number of bins into which the light curve
was partitioned.

\begin{table*}[t]
\centering
\caption{Best-fit parameters for the variability power spectra of GRS 1739-278.}
\label{tbl-powspec}
\hspace*{-1cm}\begin{tabular}{cc|cccc|cc|c}

$ID$ & Band$^a$                                             & $\delta f_{qpo}$,  Hz $^b$   & $f_{qpo}$,  Hz  $^c$             & $\delta f_{zl}$,  Hz  $^d$   & $Q=\frac{f_{qpo}}{\delta f_{qpo}}$ $^e$ & $rms_{qpo}, \%$ $^f$      & $rms_{tot}, \%$ $^f$         &$2\Delta L$ $^g$( $log(p)$ $^h$) \\ \hline
\multirow{3}{*}{02}                                     & \begin{tabular}[c]{@{}c@{}}F\\ \end{tabular} & $0.03\pm0.01$          & $0.106\pm0.004$            & $0.33\pm0.03$          & $3.4\pm1.3$                        & $13.9\pm1.8$         & $30.0\pm1.0$        &    $39$($-7.8$)                 \\
                                                        & \begin{tabular}[c]{@{}c@{}}A\\  \end{tabular}  & $0.04\pm0.02$          & $0.103\pm0.005$            & $0.32_{-0.05}^{+0.04}$ & $8.0\pm4.2$                        & $13.5_{-2.5}^{+2.3}$ & $27.0\pm1.0$        &  $28$($-5.4$)                 \\
                                                        & \begin{tabular}[c]{@{}c@{}}B\\  \end{tabular}   & $0.03_{-0.02}^{+0.01}$ & $0.113_{-0.012}^{+0.004}$ & $0.31_{-0.04}^{+0.03}$ & $4.2\pm2.5$                        & $14.5_{-2.7}^{+2.4}$ & $32.0\pm1.0$        &         $27$($-5.2$)                \\ \hline
\multirow{3}{*}{03}                                     & F                                                      & $0.03\pm0.01$          & $0.377_{-0.003}^{+0.004}$  & $0.62_{-0.04}^{+0.05}$ & $11.7\pm3.4$                       & $9.9_{-1.0}^{+0.9}$  & $27.0\pm1.0$        &   $81$($-16.7$)                  \\
                                                        & A                                                      & $0.03\pm0.01$          & $0.379\pm0.005$            & $0.57\pm0.09$          & $11.6\pm5.1$                       & $8.7_{-1.3}^{+1.2}$  & $23.0\pm1.0$                                                  & $37$($-7.3$)                   \\
                                                        & B                                                      & $0.04\pm0.01$          & $0.377_{-0.004}^{+0.005}$  & $0.61\pm0.05$          & $10.3\pm3.6$                       & $11.2\pm1.3$         & $31.0\pm1.0$                                                      & $58$( $-11.8$)              \\ \hline
\multirow{3}{*}{04}                                     & F                                                      & $0.29\pm0.03$          & $2.182\pm0.012$            & $1.59\pm0.28$          & $7.6\pm0.9$                        & $10.3\pm0.4$         & $14.0\pm1.0$                                                            & $338$($-72.2$)              \\
                                                        & A                                                      & $0.32_{-0.11}^{+0.1}$  & $2.208\pm0.039$            & $1.47_{-0.84}^{+0.64}$ & $6.8\pm2.3$                        & $7.2_{-0.9}^{+0.8}$  & $10.0\pm1.0$                                                            & $45$($-9.0$)                \\
                                                        & B                                                      & $0.21\pm0.03$          & $2.19\pm0.01$              & $5.09_{-0.67}^{+0.64}$ & $10.4\pm1.5$                       & $13.5\pm0.6$         & $25.0\pm1.0$                                                                 & $295$($-62.9$) \\ \hline
\multirow{3}{*}{05}                                     & F                                                      & $0.34\pm0.04$          & $1.69\pm0.01$              & $0.89_{-0.20}^{+0.19}$  & $5.0\pm0.6$                        & $12.6\pm0.5$         & $15.0\pm0.0$                                                                 & $262$($-55.8$)                \\
                                                        & A                                                      & $0.31_{-0.13}^{+0.11}$ & $1.73\pm0.04$              & $1.08_{-0.66}^{+0.44}$ & $5.6\pm2.3$                        & $7.7\pm1.1$          & $10.0\pm1.0$                                                   & $31$($-6.1$)                 \\
                                                        & B                                                      & $0.24_{-0.03}^{+0.04}$ & $1.7\pm0.01$               & $5.17_{-0.77}^{+0.83}$ & $7.2\pm1.1$                        & $15.5\pm0.8$         & $27.0\pm1.0$                                                   & $235$( $-49.7$)                 \\ \hline
\multirow{3}{*}{06}                                     & F                                                      & $0.4_{-0.17}^{+0.15}$  & $5.07\pm0.06$              & $1.5_{-0.34}^{+0.32}$  & $12.7\pm5.3$                       & $6.7\pm0.9$          & $12.0\pm1.0$        &               $32$($-6.3$)                  \\
                                                        & A                                                      & -                      & -                          & $0.75_{-0.52}^{+0.21}$ & - & -                    & $6.0_{-2.0}^{+1.0}$ &  $0(-)$                    \\
                                                        & B                                                      & $0.24\pm0.10$           & $5.06\pm0.03$              & $2.18_{-0.48}^{+0.49}$ & $21.1\pm9.1$                       & $8.9\pm1.2$          & $19.0\pm1.0$        &                   $34$($-6.7$)                  \\ \hline
\multirow{3}{*}{07}                                     & F                                                      & $0.48\pm0.08$          & $2.48\pm0.03$              & $0.93_{-0.19}^{+0.18}$ & $5.2\pm0.9$                        & $10.6\pm0.6$         & $15.0\pm1.0$                                               & $120$($-25.1$)                  \\
                                                        & A                                                      & $0.34_{-0.24}^{+0.20}$  & $2.41\pm0.09$              & $0.99_{-0.58}^{+0.35}$ & $7.1\pm4.9$                        & $6.4_{-1.7}^{+1.5}$  & $10.0\pm1.0$                                                     & $13$($-2.3$)                  \\
                                                        & B                                                      & $0.49\pm0.09$          & $2.52\pm0.03$              & $1.12_{-0.22}^{+0.20}$  & $5.1\pm1.0$                        & $15.2_{-1.0}^{+0.9}$ & $22.0\pm1.0$                                              & $90$($-18.7$)                  \\ \hline
\multirow{3}{*}{08}                                     & F                                                      & $0.54_{-0.35}^{+0.33}$ & $5.09\pm0.14$              & $1.08_{-0.38}^{+0.37}$ & $9.4\pm6.2$                        & $7.2\pm1.6$          & $12.0\pm1.0$        &   $14$($-2.5$)                   \\
                                                        & A                                                      & -                      & -                          & $2.63_{-1.44}^{+1.04}$ & - & -                    & $12.0\pm2.0$       &$0(-)$               \\
                                                        & B                                                      & -                      & -                          & $9.55_{-3.78}^{+3.34}$ & - & -                    & $21.0\pm3.0$        &  $0(-)$                 \\ \hline
\multirow{3}{*}{09}                                     & F                                                      & $0.31\pm0.06$          & $2.19\pm0.02$              & $0.28\pm0.08$          & $7.1\pm1.5$                        & $11.7\pm0.8$         & $14.0\pm1.0$        &                       $91$($-18.9$)                   \\
                                                        & A                                                      & $0.39_{-0.23}^{+0.21}$ & $2.25\pm0.09$              & $0.47_{-0.31}^{+0.10}$  & $5.8\pm3.5$                        & $7.4_{-1.5}^{+1.6}$  & $10.0\pm1.0$        &              $18$($-3.4$)                 \\
                                                        & B                                                      & $0.18\pm0.05$          & $2.15\pm0.02$              & $2.7_{-0.90}^{+0.92}$   & $11.9\pm3.1$                       & $15.0_{-1.3}^{+1.4}$ & $25.0\pm2.0$        &                                 $80$($-16.5$)                  
\end{tabular}
\caption*{a-0.5-10 (F), 0.5-3 (A), and 3-10 keV (B) energy bands; b-width of the QPO peak; c-frequency of the QPO peak; d-width of the underlying component; e-QPO quality factor; f-QPO fractional rms and total fractional rms; g-statistic of the likelihood ratio test; h-logarithm of the probability that this difference of the likelihood functions is a random variable. The errors were determined for
a 68\% confidence interval (see the text).}
\end{table*}

To find the error in the parameters of the best-fit
model, we used the Monte Carlo method. The
data were randomly selected 1000 times around the
best-fit model $P_{i,model}$ according to the law $P_i=\chi^2_{2n}(P_{i,model}/2n)$. The randomly selected data were
also fitted by the model using the maximum likelihood
method. The sought-for error in a parameter was
found as a difference between the mean value and the
lower (upper) limit on the parameter corresponding to
the 16\% (84\%) quantile of the distribution.

\begin{figure*}

\centering
\includegraphics[width=0.75\textwidth]{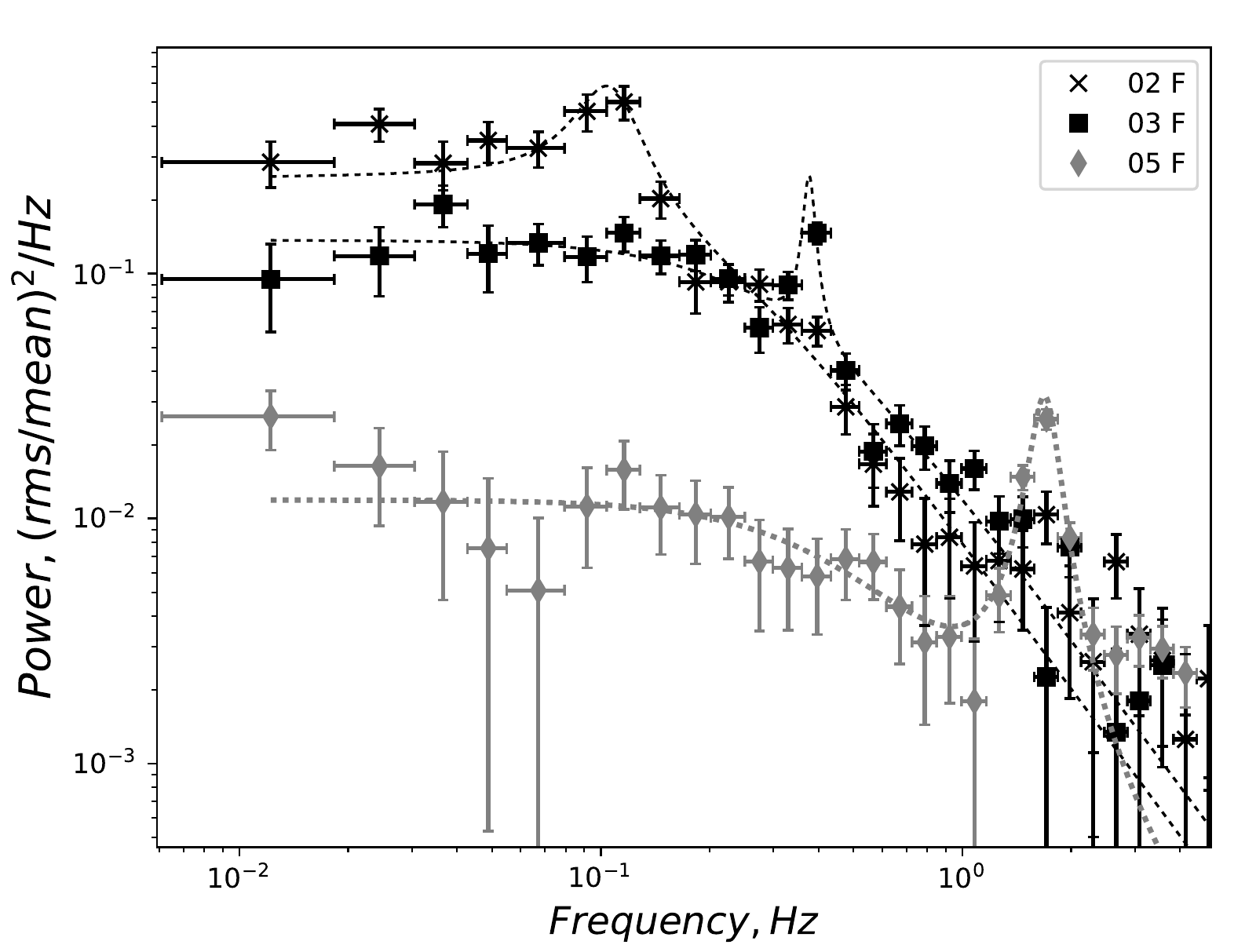}
\caption{Variability power spectra for the source in percentage normalization for observations 02 (black crosses), 03 (black squares), and 05 (gray diamonds) in the F (0.5-10 keV) band. The thin lines indicate the model fits to the spectra. The Poisson noise level was subtracted. }
\label{fig-pds1}
\end{figure*}

\begin{figure*}
\centering
\includegraphics[width=0.75\textwidth]{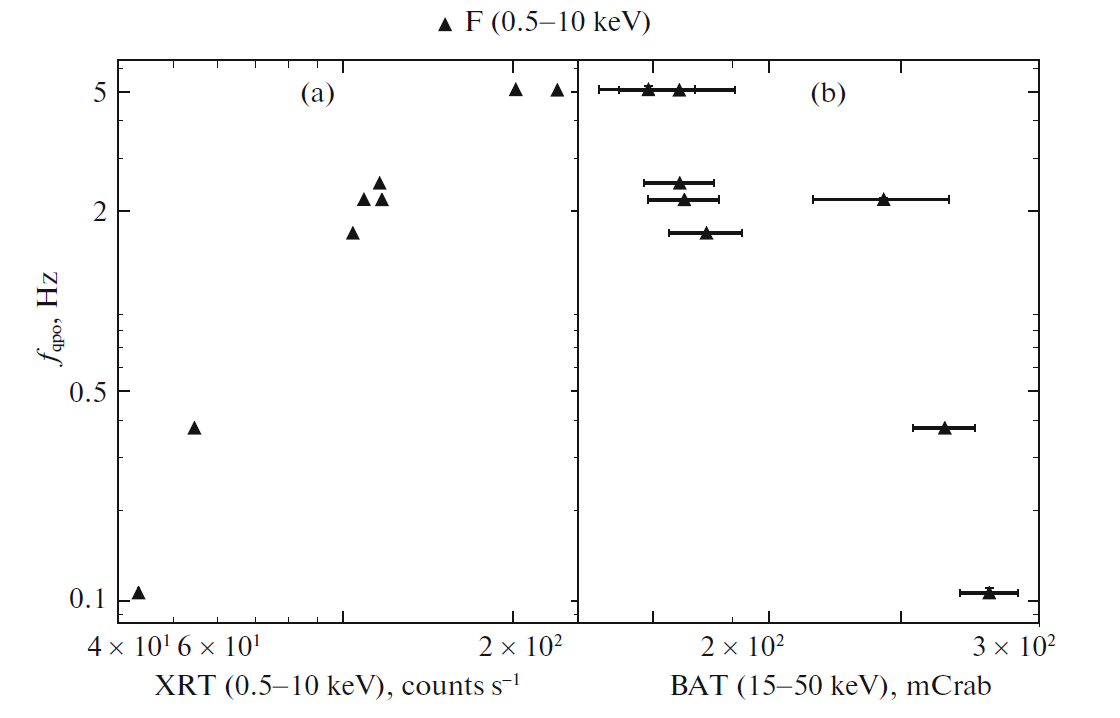}
\caption{(a) QPO frequency versus 0.8-10 keV flux. (b) QPO frequency versus 15-50 keV flux.}
\label{fig-timing_params1}
\end{figure*}

To determine the QPO significance, we calculated
the doubled difference of logarithmic likelihood functions
$2 \log(L_{qpo}/L_{null})$, where  $L_{qpo}$ and  $L_{null}$ are the
values of the likelihood function for the model with
and without QPOs, respectively, and the probability
that this difference is a random variable. The difference
of the likelihood functions has a $\chi^2_k$ distribution
(Cash 1979), where k is the difference of the numbers
of free parameters in the models with and without
QPOs, in our case, k = 3.

The results of fitting the power spectra with QPOs
are presented in Table 5. No QPOs were recorded in
observation 06 in the A band and in observation 08 in
both А and В. For these power spectra we calculated
an upper limit on the QPO fractional rms at 90$\%$ confidence
by assuming the QPO frequency and quality
factor in the A and B band to coincide with those in
the F band. For observation 06 in the A band the
upper limit is $r_{qpo}~<~6\%$ ; in observation 08, $r_{qpo}~<~10$
and 15$\%$ for the A and B bands, respectively.
It can be seen from Table 5 that the QPO frequency
does not depend on the energy band. Note
that, as follows from the literature, the systems
with black hole candidates show both no correlation
between the QPO frequency and energy and direct
and inverse proportionality (Yan et al. 2012; Li
et al. 2013a, 2013b).

Figure 3 presents the power spectra with QPOs in
the full energy band (0.5-10 keV) for several observations
(02, 03, and 05). The QPO frequency is clearly
seen to change from observation to observation.

To determine the type of QPOs, it is necessary to
measure the parameters of both the QPO peak itself
and the broadband noise. It follows from Fig. 3 and
Table 5 that broadband noise whose total fractional
rms is greater than 10$\%$ is present in the power spectra
under study at low frequencies, which, as was
said in the Introduction, is characteristic for type-
C QPOs. We constructed the dependence of the
QPO frequency on the flux in the soft (0.5-10 keV)
and hard (15-50 keV) energy bands (Fig. 4). Since
the contribution of the disk component in the 15-
50 keV energy band is minor, this may be considered
as the dependence of the QPO frequency on the flux
in the power-law component. It can be seen from the
figure that the dependence of the QPO frequency on
the flux in the soft and hard energy bands is direct
and inverse, respectively; such a behavior is also
characteristic for type-C QPOs (Motta et al. 2011).
Stiele et al. (2011) showed that type-B QPOs are
observed only at certain photon indices of the spectral
component describing the Comptonized radiation. At
the transition stage from the hard to soft state the
photon index must be greater than or of the order
of 2.2. In our case, it can be seen from Table 4 that
the photon index is less than or of the order of 2, which
again provides evidence for type-C QPOs.

For several systems, it was shown on the basis of
Fourier spectroscopy that the corona makes a major
contribution to the system's variability (Churazov
et al. 2001; Sobolewska and Zycki 2006), i.e., the
fractional rms must decrease with decreasing contribution
of the power-law component to the flux,
which we observe. In observations 02 and 03, when
the fractional rms in the A and B energy bands is
determined by the power-law component, the total
fractional rms in the soft energy band (A) is smaller
than that in the hard (B) energy band by a factor
of 1.2-1.3. At the same time, in observations 04-09, when the disk component is also present in the
soft energy band, the fractional rms in the A band is
smaller than that in the B band by a factor of 2-3.

For observations 10-42, when no QPOs were
recorded, we determined the total fractional rms in
the 0.5-10 keV energy band. For this purpose, we
fitted the power spectra either by a power law with
a constant or only by a constant by the maximum
likelihood method. If we failed to extract the powerlaw
component, then the source's limiting white noise
power was estimated. For this purpose, we searched
for $P_{source}$ at which the measured power in the frequency
range 0.01-50 Hz was the 10\% quantile of
the normal distribution $\mathcal{N}(P_{noise}+P_{source}, (P_{noise}+P_{source})/Nn)$, where N is the number of frequency
bins and $P_{source}$  is the measured Poisson noise level.

A diagram of the derived dependence of the total
fractional rms on hardness (the ratio of the 4-10 and 0.5-4 keV fluxes) is shown in Fig. 2b. It follows from
the figure that the total fractional rms of the source in
the time of observations decreased from $\sim$30 to $\sim$8\%
or less.

\subsection{Observed States during the 2014 Outburst}

It follows from our spectral analysis (see Tables 1
and 4) that during observations 01-03 the photon
index of the power-law component was $\sim 1.2 - 1.4$,
while it follows from Table 5 that the total fractional
rms during observations 02 and 03 was 30 and 27\%,
respectively. The values of these parameters suggest
that the system was in the low/hard state in the period
from observation 01 to 03.

From observation 04 to 09 the total fractional rms
decreased to 12-15\%, type-C QPOs were observed
in the variability power spectrum, and the accretion
disk contribution, along with the power-law component,
is recorded in the source's energy spectrum,
which is typical for the hard intermediate state. The
total fractional rms was $10\pm2$ \% during observation
10, $8\pm3\%$ in observation 11, and 14-16\% in
observations 12 and 13, which also provides evidence
for the hard intermediate state.

From observation 14 to 23 the source's energy
spectrum is still described by the model of an accretion
disk with a power-law component, but the
fractional rms of the source dropped below 10\%. In
many observations we managed to obtain only upper
limits at a level <10\%. Thus, it can be concluded
that the system passed to the soft intermediate state
between observations 13 and 14.

During observation 11 (when not only the total
fractional rms, but also the contribution of the powerlaw
component to the flux from the system decreased
almost by a factor of 2 compared to the adjacent observations,
see Table 1) the system may have passed
to the soft intermediate state, but this cannot be
asserted based on the available data. Despite the
fact that fitting the data by the phabs*(diskbb +powerlaw) and phabs*diskbb models gives identical
$\chi^2$ values, we think that the first model is most probable,
because the source exhibits a significant flux
(150 mCrab) in the 15-50 keV energy band.

From observation 24 to 42 the source's energy
spectra are well described by the model of a multitemperature
disk with low-energy absorption, i.e., it
can be argued that the system passed to the high/soft
state. Although adding the power-law component
when fitting some of the spectra formally reduces the
$\chi^2$ value, nevertheless, first, the contribution of the
power-law component is minor ($≤30\%$) and, second,
a weak power-law can also be observed in the
high/soft state (Belloni and Motta 2016). It is worth
noting that the powerlaw model has no low-energy
cutoff and the estimate of the contribution from the
power-law component to the total flux is an upper
limit, i.e., the fraction of the nonthermal component
is actually smaller. Furthermore, it can be seen from
Fig. 1 that after observation 23 the source is barely
recorded in the 15-50 keV energy band, which also
provides evidence for the transition to the high/soft
state.

The source's characteristic spectra corresponding
to the low/hard, intermediate low/hard, and high/soft
states (observations 01, 09, and 42, respectively) are
presented in Fig. 5.

\begin{figure*}[t]

\centering
\includegraphics[width=\textwidth]{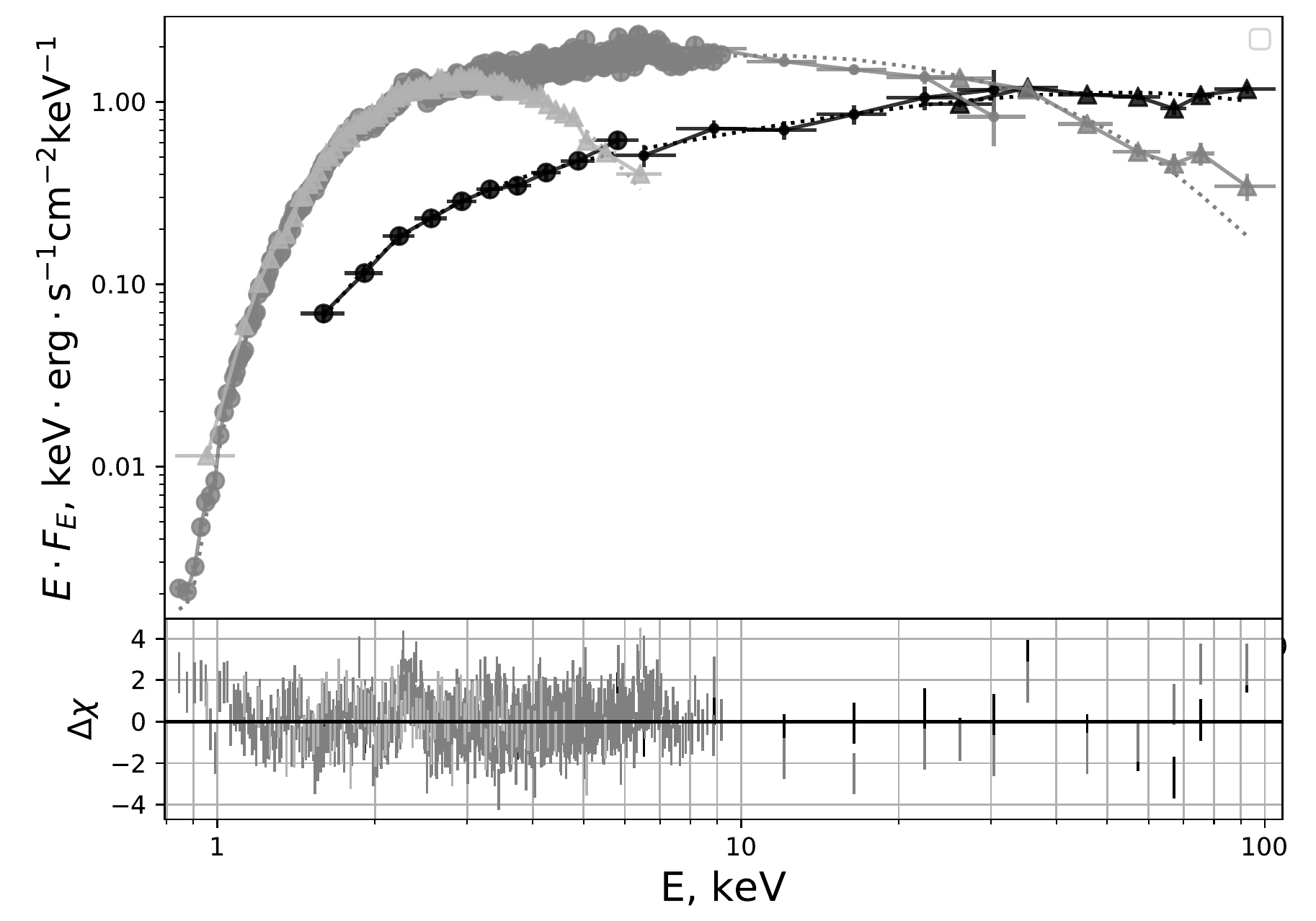}
\caption{Characteristic energy spectra of GRS 1739-278 and their best-fit-models during different spectral states: observation 01 (low/hard state)-black circles, dots, and triangles: the Swift/XRT, INTEGRAL/JEMX, and INTEGRAL/ISGRI data, respectively; observation 09 (intermediate state)-dark-gray circles, dots, and triangles: the Swift/XRT, INTEGRAL/JEMX, and INTEGRAL/ISGRI data, respectively; observation 42 (high/soft state)-light-gray triangles (Swift/XRT data). The thin
dash-dotted lines indicate the best-fit models (see Tables 1 and 3). The lower panel shows the deviation of the data from the models.}
\label{fig-spe}
\end{figure*}

Figure 6 shows the periods of time and the states
in which the source was and presents the time dependences
of the accretion disk temperature at the inner
radius, the contribution of the power-law component
to the total flux, and the total fractional rms. The
accretion disk temperature at the inner radius and the
contribution of the power-law component were taken
from the model in which the photon index was fixed
at 2.

\begin{figure*}[t]
\centering
\includegraphics[width=\textwidth]{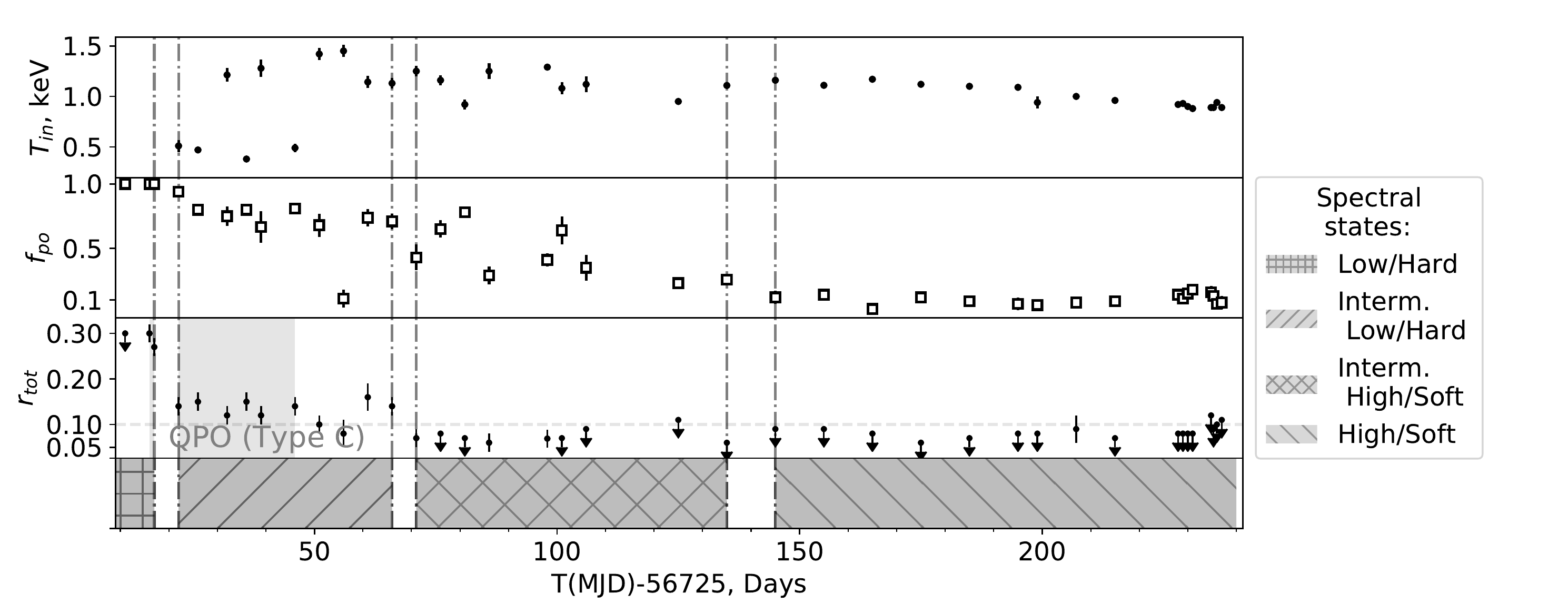}
\caption{(a) Accretion disk temperature at the inner radius versus time. (b) Contribution of the power-law component to the total flux versus time. (c) Total fractional rms versus time. The accretion disk temperature at the inner radius and the contribution of the power-law component are given for the model in which the photon index was fixed at 2. The periods of time when the source was in the low/hard, intermediate hard and soft, and high/soft states are also marked in the figure.}
\label{fig-states}
\end{figure*}

On the whole, our results on the transitions between
states are consistent with those from Yan and
Yu (2017) and Wang et al. (2018).

\subsection{Mini-Outbursts of the System}

We performed an analysis of the light curve for
GRS 1739-278 over the entire period of observations
since its discovery aimed at searching for undetected
outbursts. From 1996 to mid-2011 the source
was regularly observed by the ASM/RXTE all-sky
monitor in the 1.2-12 keV energy band (Levine
et al. 1996). According to these data, after the bright
1996 outburst the source exhibited no outburst activity
and was not detected. Since 2005 the source has
been observed almost continuously by the Swift/BAT
telescope in the 15-50 keV energy band. From 2005
to 2014 no outbursts were detected on the light curve
and the mean flux was $1.0\pm0.4$ mCrab. After the end
of the 2014 outburst the mean flux from the source
rose to $9.7 \pm 0.2$ mCrab.

\begin{figure*}

\centering
\includegraphics[width=\textwidth]{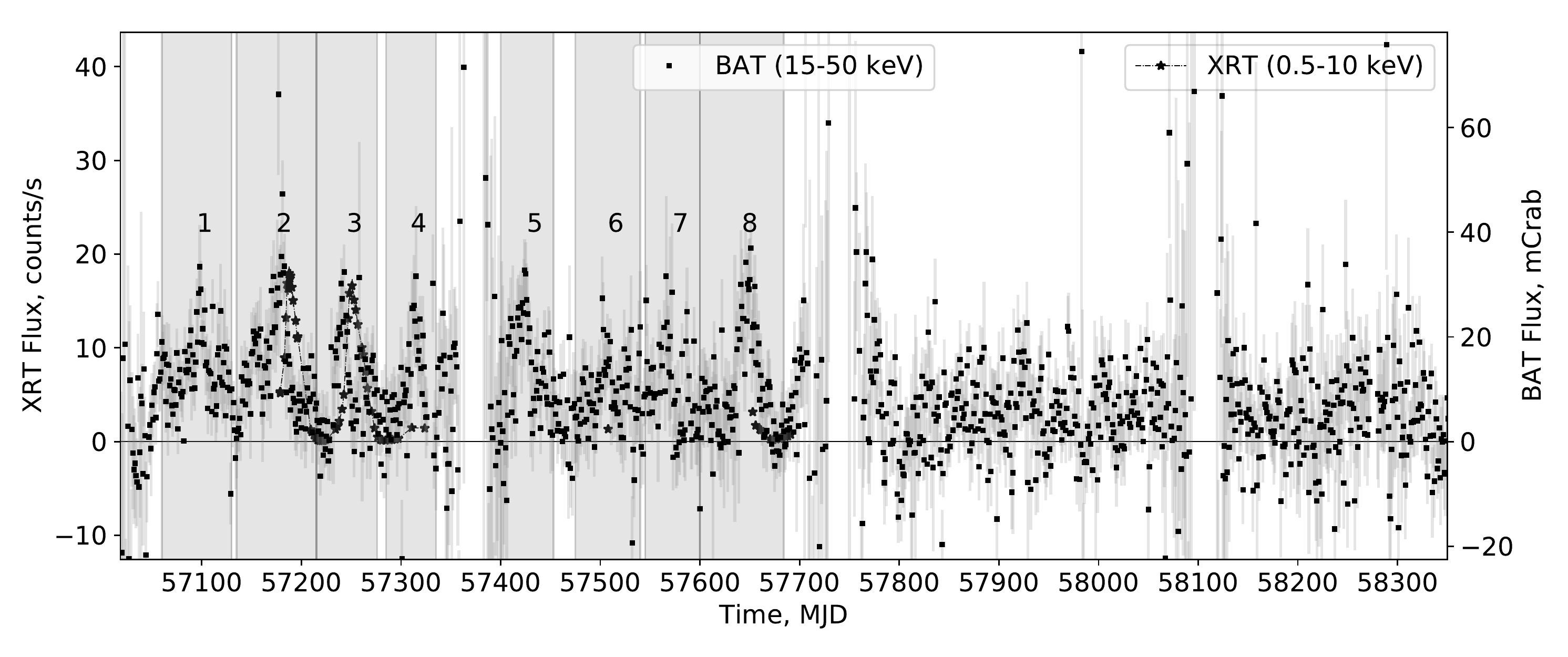}
\caption{Light curve of GRS 1739-278 for 2015-2016 from Swift/XRT (black stars, the data were averaged over one observation) and Swift/BAT (black squares, the data were averaged over one day) data. The shaded regions indicate the time intervals during which the outburst significance was determined (see the text and Fig. 8)}
\label{fig-lc2}
\end{figure*}

\begin{figure*}
\centering
\includegraphics[width=\textwidth]{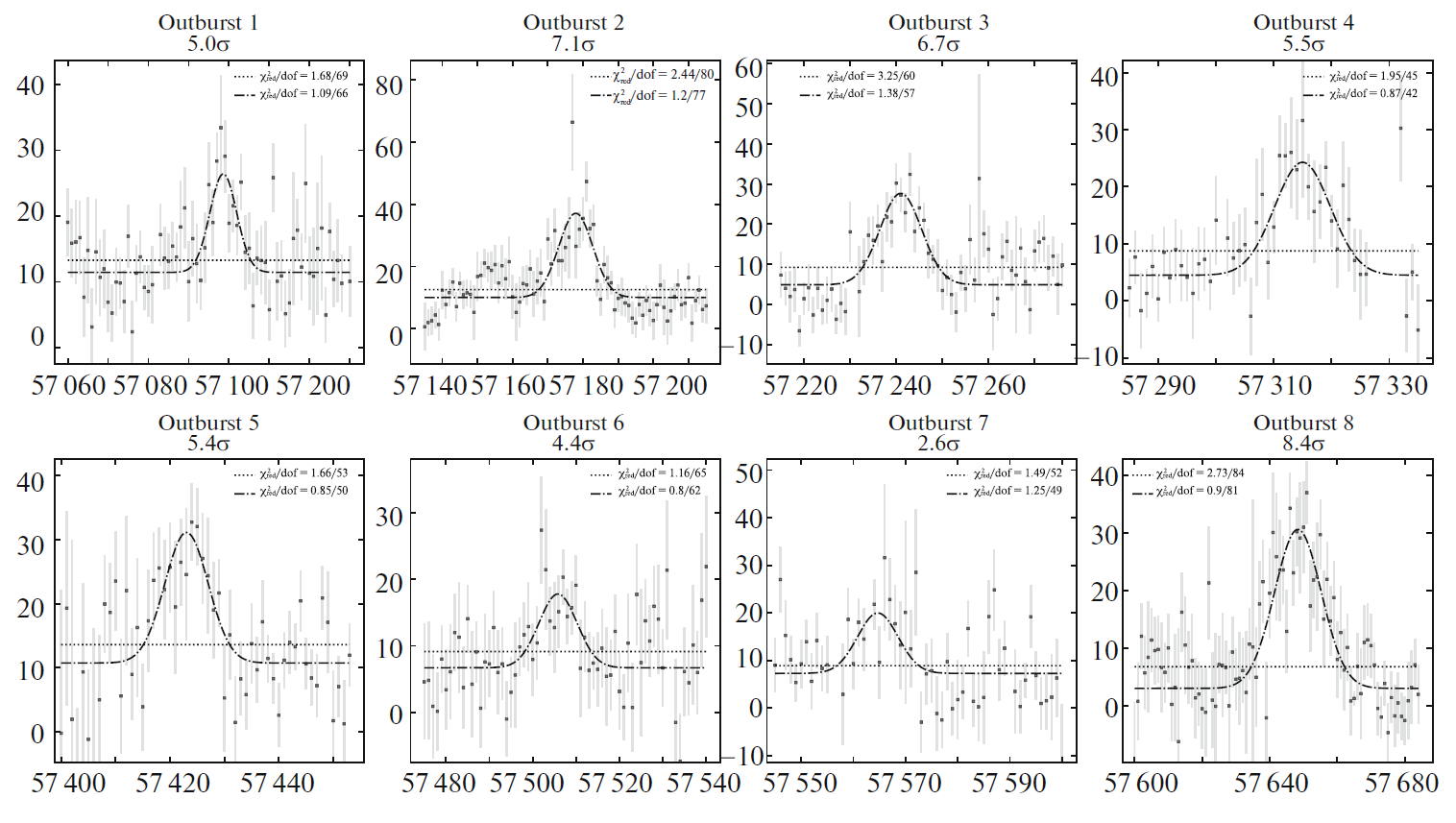}
\caption{Fitting the source's outbursts detected from Swift/BAT data (15-50 keV) (see Fig. 7) by a constant (black dotted line) and a constant with the addition of a Gaussian profile (black dash-dotted line). The $\chi^2$ value is given for the fit by each model. The outburst detection significance is specified in the panel header.}
\label{fig-bat-flares}
\end{figure*}

\begin{figure*}

\centering
\includegraphics[width=0.75\textwidth]{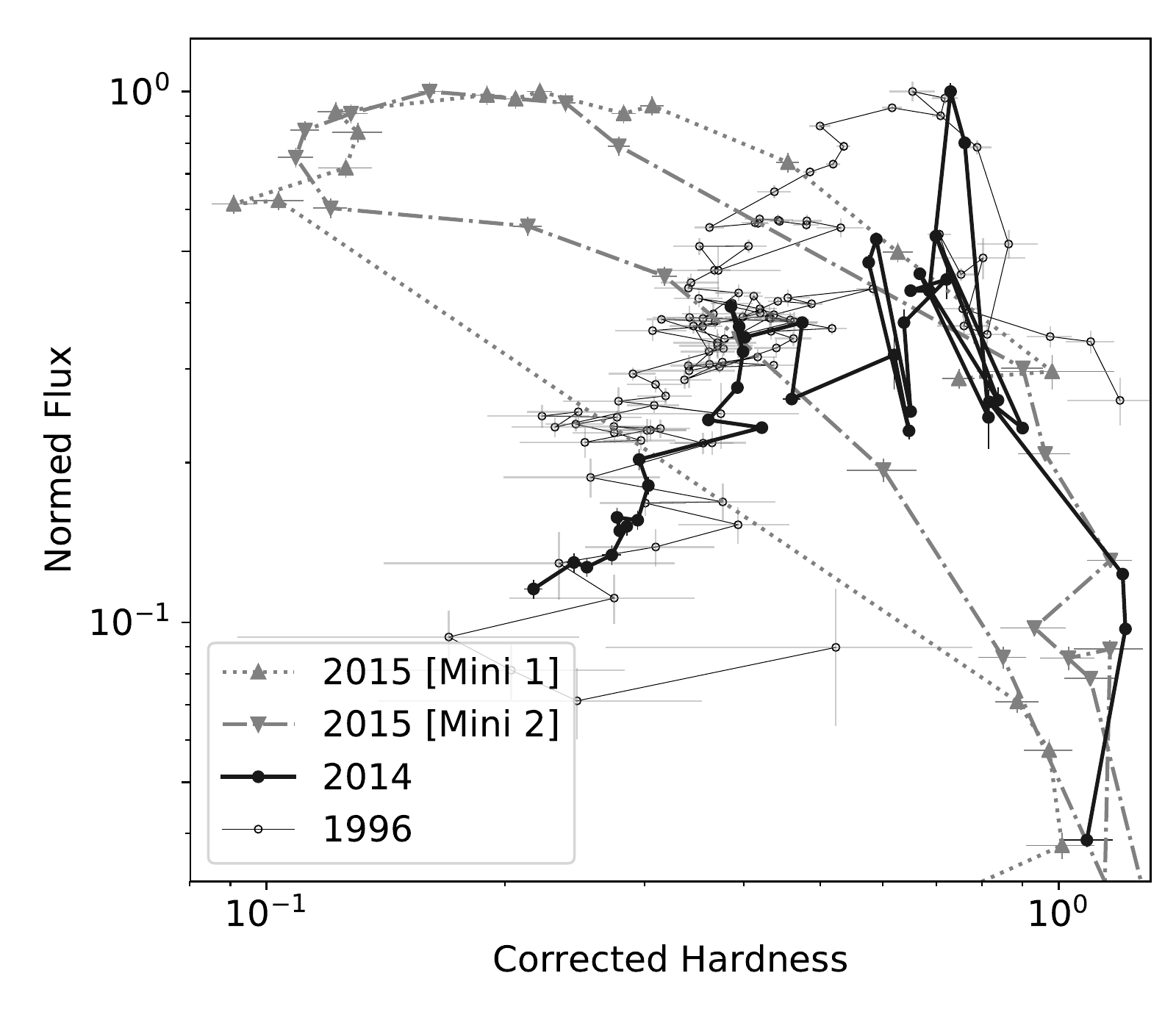}
\caption{Hardness-intensity diagram for the 1996 and 2014 outbursts as well as the first and second 2015 mini-outbursts (during which the system exhibited the transition from the low/hard to high/soft state), marked by the white circles, black circles, gray upper and lower triangles, respectively. The corrected hardness (see the text) is along the horizontal axis, the counts normalized to the maximum of each outburst is along the vertical axis: based on the ASM and XRT data in the 1.2-12 and 0.5-10 keV energy bands, respectively.}
\label{fig-hid2}
\end{figure*}

\begin{figure*}
\centering
\includegraphics[width=0.5\textwidth]{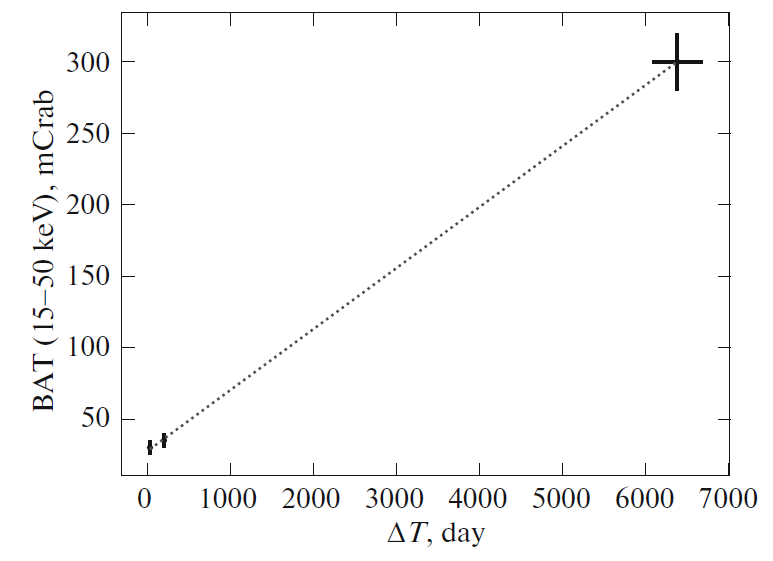}
\caption{BAT flux (in mCrab) at the peak of the outburst low/hard state versus time elapsed from the maximumin the low/hard state in the previous outburst. The dotted line indicates a linear best-fit model in the form 0.043 $mCrab/day$ $\Delta T$ + 27 mCrab. }
\label{fig-deltat}
\end{figure*}

A detailed analysis of the Swift/BAT light curve
after the 2014 outburst showed that, apart from the
mini-outbursts mentioned in the literature, the system
exhibited several more similar events. To determine
the statistical significance of the detected
outbursts, we performed an analysis in which we
partitioned the light curves into bins containing these
outbursts (indicated by the gray rectangles in Fig. 7)
and fitted the time dependence of the flux by two
models: a constant and a constant with a Gaussian
profile as the first approximation for the outburst profile.
The outburst detection significance was defined
as the probability of the difference of the $\chi^2$ values
for both best-fit models (F-test). The results of our
analysis are shown in Fig. 8; the outburst detection
significance is given above each panel. It follows from the figure that the mini-outbursts mentioned in the
literature had a significance of 7-8$\sigma$ (mini-outbursts
2, 3, and 8), while the detected four mini-outbursts
have a significance of 4-5.5$\sigma$. Since outburst 7 has
a low significance, 2.6$\sigma$, we did not include it in our
final conclusions. After the failed 2016 outburst the
system returned to a quiescent state with a mean flux
of $5.0 \pm 0.3$ mCrab.

\subsection{Evolution of the Outbursts of the System in 1996,
2014, and 2015}

Using the RXTE/ASM archival data, we constructed
the hardness-intensity diagram for the 1996
outburst and compared it with that for the 2014 outburst
and the 2015 mini-outbursts, when the system
passed to the high/soft state. To make a proper
comparison of the diagrams, it is necessary to take
into account the difference in the energy bands and
characteristics of the instruments. For this purpose,
we calculated the hardness using the best-fit spectral
models for several states. In the 1996 outburst
we chose the KVANT/TTM observations on February
6-7, 1996, and the RXTE/PCA observations on
March 31, 1996, and May 29, 1996 (see Borozdin
et al. 1998). For the 2014 outburst we used observations
01, 11, and 32. The first observations for
each outburst correspond to the time of the largest
recorded hardness, the second observations refer to
the time of the source's maximum soft X-ray intensity,
and the third observations refer to the high/soft
state, when the power-law contribution to the total
flux is minor. Using this sample of observations, we
calculated the hardness for the 5-10 and 1.5-5 keV
energy bands. The hardness ratios were found to
be 1.29, 0.57, and 0.30, respectively, in the 1996 outburst
and 1.29, 0.64, and 0.23 in the 2014 outburst.
Having calculated the hardness from the light curves
in the above reference observations, we found the ratios
of the "true" $\,$ (based on the models) and observed
(based on the light curves) hardnesses. The scatter
of these ratios relative to the mean value is $12-15\%$;
therefore, we used the mean value as a coefficient to
convert the observed hardness-intensity diagram to
the true one. The same coefficient was used For the
2014 outburst and the 2015 mini-outbursts. For the
convenience of comparing the diagram shapes, we
also normalized the flux to its maximum value. The
derived diagrams are presented in Fig. 9. It can be
seen from the figure that during the bright outbursts
the curves have a similar shape, with the behavior of
the source on the hardness-intensity diagram during
the bright outbursts differing significantly from its
behavior during the mini-outbursts. We estimated
the luminosity at which a minimum hardness was
reached during the outbursts by assuming the distance
to the system to be 8.5 kpc: $L_{1.2-12 \text{\,keV}}\sim 1.5\times 10^{37}$ erg s$^{-1}$ for the 1996 outburst, $L_{0.5-10 \text{\,keV}}\sim 2\times 10^{37}$ erg s$^{-1}$ for the 2014 outburst, and$L_{0.5-10 \text{\,keV}}\sim(5-6)\times 10^{36}$ erg s$^{-1}$ for the mini-outbursts.

Yu et al. (2007) and Wu et al. (2010) constructed
the dependence of the peak flux in the 20-160 keV
energy band during the low/hard state on the time
between the current and previous peak fluxes in the
low/hard state for GX 339-4 (a low-mass binary
system with a black hole candidate) and attempted
to fit this dependence by a linear law. We constructed
the same dependence of the peak flux in
the 15-50 keV energy band in the low/hard state on the time to the previous peak in the low/hard
state for GRS 1739-278 by taking into account the
bright 1996 and 2014 outbursts and the 2015 mini-outbursts
(Fig. 10). The time of the transition to
the hard state at the end of the 2014 outburst was
taken from Wang et al. (2018). The linear dependence
that best fits the data for GRS 1739-278 looks
as follows: $F_{hard}(\Delta T)=(0.043\pm 0.003) \frac{\text{mCrab}}{\text{day}}\Delta T+(27\pm 2 ) \text{mCrab}$.

\section{CONCLUSIONS}

In this paper we performed a joint study of the
spectral and temporal evolution of GRS 1739-278
during the 2014 outburst and made a comparative
analysis of the system's behavior during the remaining
outbursts mentioned in the literature and in the
periods between them. Our results can be briefly
summarized as follows.

\begin{itemize}

\item  We showed that during the 2014 outburst the
system passed to the hard intermediate state
22 days after the outburst onset, to the soft
intermediate state 66 days later (possibly exhibiting
this state on day 55 and returning to
the hard intermediate state no later than 4 days
after), and to the high/soft state 145 days later.

\item  QPOs in the frequency range 0.1-5 Hz were
detected during the outburst of GRS 1739-278
in 2014. All QPOs are type-C ones. No energy
dependence of the QPO frequency was found.

\item We showed that after the 2014 outburst the
system passed to the regime of mini-outburst
activity and, apart from the three mini-outbursts
mentioned in the literature (Yu and
Yan 2017; Mereminskiy et al. 2017), we
detected four more mini-outbursts with a comparable
($\sim$20 mCrab) flux in the hard energy
band (15-50 keV).

\item  We showed that the hardness-intensity diagram
for the 2015 mini-outbursts, during
which the system exhibited the transition to
the high/soft state, differs from that for the
bright 1996 and 2014 outbursts: the minimum
hardness during the mini-outbursts was
reached at fluxes of at least $60-80\%$ of the peak
one, while in the bright outbursts the minimum
hardness was reached at fluxes of $\sim10\%$ of the
peak one. The 0.5-10 keV luminosity of the
source corresponding to these times differed
approximately by a factor of 3:  $L_{0.5-10\text{\, keV}}\sim 2\times 10^{37}$ erg s$^{-1}$ for the bright outburst and
$L_{0.5-10\text{\,keV}}\sim(5-6)\times 10^{36}$ erg s$^{-1}$ for the
mini-outbursts.

\item We constructed the dependence of the peak
flux in the hard energy band during the low/hard
state on the time interval between outbursts.
This dependence can be fitted by a linear law,
which may point to the dependence of the
system's peak flux in the low/hard state on the
mass of the accretion disk being accumulated.
\end{itemize}

\section{ACKNOWLEDGMENTS}

This work was financially supported by RSF grant
no. 14-12-01287. We used the data provided by the
UK Swift Science Data Center at the University of
Leicester and the INTEGRAL Science Data Centers
at the University of Geneva and the Space Research
Institute of the Russian Academy of Sciences.

\section{REFERENCES}

\begin{itemize}
\itemsep0em 
\item[] 1. K. A. Arnaud, Astron. Data Anal. Software Syst. V
\textbf{101}, 17 (1996).
\item[]2. T. M. Belloni, Lect. Notes Phys. \textbf{794}, 53 (2010).
\item[]3. T. M. Belloni and S. E. Motta, Astrophys. Space Sci.
Lib. \textbf{440}, 61 (2016).
\item[]4. K. N. Borozdin and S. P. Trudolyubov, Astrophys.
J. \textbf{533}, L131 (2000).
\item[]5. K. Borozdin, N. Alexandrovich, R. Sunyaev, et al.,
IAU Circ. \textbf{6350} (1996).
\item[]6. K.N. Borozdin, M. G. Revnivtsev, S. P. Trudolyubov,
et al., Astron. Lett. \textbf{24}, 435 (1998).
\item[]7. D. N. Burrows, J. E. Hill, J. A. Nousek, et al., Space
Sci. Rev. \textbf{120}, 165 (2005).
\item[]8. F. Capitanio, T. Belloni, M. del Santo, et al., Mon.
Not. R. Astron. Soc. \textbf{398}, 1194 (2009).
\item[]9. W. Cash, Astrophys. J. \textbf{228}, 939 (1979).
\item[]10. P. Durouchoux, I. A. Smith, K. Hurley et al., IAU
Circ. \textbf{6383}, 1 (1996).
\item[]11. P. A. Evans, A. P. Beardmore, K. L. Page, et al.,
Astron. Astrophys. \textbf{469}, 379 (2007).
\item[]12. C. Ferrigno, E. Bozzo, M. del Santo, et al., Astron.
Astrophys. \textbf{537}, L7 (2012).
\item[]13. E. Filippova, E. Bozzo, and C. Ferrigno, Astron. Astrophys.
\textbf{563}, A124 (2014a).
\item[]14. E. Filippova, E. Kuulkers, N.M. Skadt, et al., Astron.
Telegram \textbf{5991}, 1 (2014b).
\item[]15. F. Furst, M. A. Nowak, J. A. Tomsick, et al., Astrophys.
J. \textbf{808}, 122 (2015).
\item[]16. M. R. Gilfanov, Lect. Notes Phys. \textbf{794}, 17 (2010).
\item[]17. S. Grebenev, R. Sunyaev,M. Pavlinsky, et al., Astron.
Astrophys. Suppl. Ser. \textbf{97}, 281 (1993).
\item[]18. S. Grebenev, R. Sunyaev, and M. Pavlinsky, Adv.
Space Res. \textbf{19}, 15 (1997).
\item[]19. J. Homan, R. Wijnands,M. van der Klis, et al., Astrophys.
J. \textbf{132}, 377 (2001).
\item[]20. A. Ingram, C. Done, and P. C. Fragile,Mon. Not. R.
Astron. Soc. \textbf{397}, L101 (2009).
\item[]21. M. van der Klis, in Timing Neutron Stars, Ed. by
H. Ogelman and E. P. J. van den Heuvel, NATO ASI
Ser. C \textbf{262}, 27 (1988).
\item[]22. H. A. Krimm, S. T. Holland, R. H. D. Corbet, et al.,
Astrophys. J. \textbf{209}, 14 (2013).
\item[]23. H. A. Krimm, S. D. Barthelmy, W. Baumgartner,
et al., Astron. Telegram \textbf{5986}, 1 (2014).
\item[]24. D. A. Leahy,W. Darbro,R. F. Elsner, et al., Astrophys.
J. \textbf{266}, 160 (1983).
\item[]25. A.M. Levine, H. Bradt, W. Cui, et al., Astrophys. J.
\textbf{469}, L33 (1996).
\item[]26. Z. B. Li, J. L. Qu, L.M. Song, et al., Mon. Not. R.
Astron. Soc. \textbf{428}, 1704 (2013a).
\item[]27. Z. B. Li, S. Zhang, J. L. Qu, et al., Mon. Not. R.
Astron. Soc. \textbf{433}, 412 (2013b).

\item[]28. I. Mereminskiy, R. Krivonos, S. Grebenev, et al., Astron.
Telegram \textbf{9517}, 1 (2016).
\item[]29. I. A. Mereminskiy, E. V. Filippova, R. A. Krivonos,
S. A. Grebenev, R. A. Burenin, and R. A. Sunyaev,
Astron. Lett. \textbf{43}, 167 (2017).
\item[]30. I. A. Mereminskiy, A. N. Semena, S.D. Bykov, et al.,
Mon. Not. R. Astron. Soc. \textbf{482}, 1392 (2019).
\item[]31. J. M. Miller, J. A. Tomsick, M. Bachetti, et al., Astrophys.
J. \textbf{799}, L6 (2015).
\item[]32. S.Motta, T.Munoz-Darias, and T.Belloni,Mon.Not.
R. Astron. Soc. \textbf{408}, 1796 (2010).
\item[]33. S. Motta, T. Munoz-Darias, P. Casella, et al., Mon.
Not. R. Astron. Soc. \textbf{418}, 2292 (2011).
\item[]34. J. Paul, L. Bouchet, E. Churazov, et al., IAU Circ.
\textbf{6348}, 1 (1996).
\item[]35. R. A. Remillard and J. E. McClintock, Ann. Rev.
Astron. Astrophys. \textbf{44}, 49 (2006).
\item[]36. J. Rodriguez, S. Corbel, E. Kalemci, et al., Astrophys.
J. \textbf{612}, 1018 (2004).
\item[]37. M. del Santo, T.M. Belloni, J. A. Tomsicket al.,Mon.
Not. R. Astron. Soc. \textbf{456}, 3585 (2016).
\item[]38. D. M. Smith, W. A. Heindl, C. B. Markwardt, et al.,
Astrophys. J. \textbf{554}, L41 (2001).
\item[]39. M. A. Sobolewska and P. T. Zycki, Mon. Not. R.
Astron. Soc. \textbf{370}, 405 (2006).
\item[]40. J. F. Steiner, R. Narayan, J. E. McClintock, et al.,
Publ. Astron. Soc. Pacif. \textbf{121}, 1279 (2009).
\item[]41. H. Stiele, S. Motta, T. Munoz-Darias, et al., Mon.
Not. R. Astron. Soc. \textbf{418}, 1746 (2011).
\item[]42. Y. Tanaka and N. Shibazaki, Ann. Rev. Astron. Astrophys.
\textbf{34}, 607 (1996).
\item[]43. M. Vargas, A. Goldwurm, J. Paul, et al., Astron.
Astrophys. \textbf{313}, 828 (1996).
\item[]44. A. Vikhlinin, E. Churazov, M. Gilfanov, et al., Astrophys.
J. \textbf{424}, 395 (1994).
\item[]45. R. Walter, R. Rohlfs, M. T. Meharga, et al., in Proceedings
of the 8th Integral Workshop on The
Restless Gamma-ray Universe INTEGRAL 2010,
PoS(INTEGRAL2010)\textbf{162}.
\item[]46. S.Wang, N. Kawai,M. Shidatsu, et al., Publ. Astron.
Soc. Jpn. \textbf{70}, 67 (2018).
\item[]47. Y. X.Wu,W.Yu, Z. Yan, et al., Astron. Astrophys. \textbf{512},
A32 (2010).
\item[]48. S. P. Yan, J. L. Qu, G. Q. Ding, et al., Astron.
Astrophys. Suppl. Ser. \textbf{337}, 137 (2012).
\item[]49. Z. Yan and W. Yu, Mon. Not. R. Astron. Soc. \textbf{470},
4298 (2017).
\item[]50. W. Yu and Z. Yan, Astrophys. J. \textbf{701}, 1940 (2009).
\item[]51. W. Yu, M. van der Klis, and R. Fender, Astrophys. J.
\textbf{611}, L121 (2004).
\item[]52. W.Yu, F. K.Lamb,R.Fender, et al., Astrophys. J. \textbf{663},
1309 (2007).
\item[]53. A. A. Zdziarski,M.Gierlinski, J.Mikolajewska, et al.,
Mon. Not. R. Astron. Soc. \textbf{351}, 791 (2004).

\end{itemize}

\end{document}